\documentclass[useAMS,usenatbib]{mn2e}

\pdfoutput=1

\usepackage{ifpdf}
\usepackage{amssymb,amsmath}
\usepackage{graphics}
\usepackage{subfigure}
\usepackage[dvips]{graphicx}
\usepackage{epstopdf}
\usepackage{textcomp}
\usepackage{hyperref}
\usepackage[english]{babel}
\usepackage{color}
\usepackage{verbatim}
\usepackage[normalem]{ulem}

\title[Stellar feedback in dwarf galaxies]{Variable interstellar radiation fields in simulated dwarf galaxies: supernovae versus photoelectric heating}

\author[Hu et al.]
{Chia-Yu Hu$^{1,2}$, Thorsten Naab$^{1}$, Simon C. O. Glover$^{3}$, Stefanie Walch$^{4}$, Paul C. Clark$^{5}$ \\
$^{1}$Max-Planck-Institut f\"ur Astrophysik,
Karl-Schwarzschild Strasse 1, D-85740 Garching, Germany\\
$^{2}$Center for Computational Astrophysics, Flatiron Institute, 162 5th Ave NY NY\\
$^{3}$Zentrum f\"ur Astronomie der Universit\"at Heidelberg, 
Institut f\"ur Theoretische Astrophysik, Albert-Ueberle-Str. 2, 69120 Heidelberg, Germany\\
$^{4}$Physikalisches Institut der Universit\"at zu K\"oln, 
Z\"ulpicher Strasse 77, D-50937 K\"oln, Germany\\
$^{5}$School of Physics and Astronomy, Cardiff University, 5 The Parade, 
Cardiff CF24 3AA, Wales, UK\\
}

\begin{document}
\maketitle

\begin{abstract}

We present high-resolution hydrodynamical simulations of isolated dwarf galaxies including self-gravity, non-equilibrium cooling and chemistry, interstellar radiation fields (IRSF) and shielding, star formation, and stellar feedback. 
This includes spatially and temporally varying photoelectric (PE) heating, photoionization, resolved supernova (SN) blast waves and metal enrichment. 
A new flexible method to sample the stellar initial mass function allows us to follow 
the contribution to the ISRF,
the metal output and the SN delay times of individual massive stars. 
We find that SNe play the dominant role in regulating the global star formation rate, shaping the multi-phase interstellar medium (ISM) and driving galactic outflows. Outflow rates (with mass-loading factors of a few) and hot gas fractions of the ISM increase with the number of SNe exploding in low-density environments where radiative energy losses are low. 
While PE heating alone can suppress star formation slightly more (a factor of a few) than SNe alone can do, 
it is unable to drive outflows and reproduce the multi-phase ISM that emerges naturally when SNe are included. 
These results are in conflict with recent results of Forbes et al. who concluded that PE heating is the dominant process suppressing star formation in dwarfs, 
about an order of magnitude more efficient than SNe.
Potential origins for this discrepancy are discussed.
In the absence of SNe and photoionization (mechanisms to disperse dense clouds), 
the impact of PE heating is highly overestimated owing to the (unrealistic) proximity of dense gas to the radiation sources. 
This leads to a substantial boost of the infrared continuum emission from the UV-irradiated dust and a far infrared line-to-continuum ratio too low compared to observations. 
Though sub-dominant in regulating star formation,
the ISRF controls the abundance of molecular hydrogen via photodissociation.

\end{abstract}

\begin{keywords}
methods: numerical,  galaxies: ISM, galaxies: evolution
\vspace{-1.0cm}
\end{keywords}

\section{Introduction}

For a long time,
stellar feedback has been an indispensable ingredient in the theory of galaxy formation (\citealp{1974MNRAS.169..229L, 1978MNRAS.183..341W, 1986ApJ...303...39D, 1999ApJ...513..142M}).
Current large-scale cosmological simulations that incorporate efficient stellar feedback have been able to successfully reproduce many properties of the observed galaxies (see reviews by \citealp{2015ARA&A..53...51S} and \citealp{2016arXiv161206891N} and references therein).
However, 
these simulations still lack the required resolutions to directly follow the feedback processes.
As such,
most of them have to resort to ad-hoc sub-grid recipes in order to make the feedback efficient.
These feedback recipes require explicit parameter calibration to certain observations and therefore should be viewed as a phenomenological ``input'' of the model,
which means that they are inadequate for investigating how stellar feedback couples with the gas dynamics on parsec scales.
To do so,
stellar feedback should be simulated from first principles without fine tuning.
This is, however, a very demanding task in terms of numerical resolution,
and the trade-off one often has to make is to focus only on a single gas cloud or a small patch of a  galaxy 
(see e.g. \citealp{2006ApJ...653.1266J, 2011MNRAS.414..321D, 2011ApJ...743...25K,
2012MNRAS.427..625W, 2012ApJ...750..104H, 2015MNRAS.454..238W, 2016ApJ...827L..29S, 2016arXiv161206891N}).

Dwarf galaxies,
thanks to their relatively small sizes and low masses,
provide a unique opportunity for simulators to conduct full-disk simulations at the required resolution \citep{2013ApJ...775..109K, 2013ApJ...779....8K, 2016MNRAS.458.3528H, 2016Natur.535..523F}.
Moreover,
they serve as ideal laboratories where numerical experiments can be conducted to gain insight on low-metallicity systems,
which are very different from local spiral galaxies and so may provide important clues to the behaviour of high-redshift galaxies which also have low metallicity.

In a previous study (\citealp{2016MNRAS.458.3528H}, hereafter Paper I) we presented a set of detailed simulations of isolated dwarf galaxies.
The model included gravity, hydrodynamics, non-equilibrium cooling and chemistry, radiation shielding, star formation, stellar feedback and metal enrichment.
We found that the star formation rate in the simulated galaxies was insensitive to the adopted (constant) strength of the interstellar radiation field (ISRF),
because the supernova (SN) feedback dominates over the photoelectric (PE) heating (which depends on the ISRF) and therefore drives the ISM out of thermal equilibrium.
These simulated galaxies have a long gas depletion time in agreement with the observed dwarf galaxies whenever SN feedback is included.

Recently,
\citet{2016Natur.535..523F} presented a set of simulations of dwarf galaxies with a similar ISM model and initial conditions at comparable resolutions,
but with one major improvement: a self-consistent ISRF directly calculated from the stellar particles.
The ISRF is computed without sophisticated and costly radiative transfer and the dust extinction is assumed to be negligible,
which should be a good approximation on the scales of interest in dust-poor dwarf galaxies.
Contrary to our findings in Paper I,
\citet{2016Natur.535..523F} concluded that PE heating is the dominant feedback process in their simulations,
and they attributed the difference between the two set of simulations to their more accurate treatment of the ISRF which varies both spatially and temporally.

In this paper,
we adopt a similar approach to the method used in \citet{2016Natur.535..523F} to compute the ISRF self-consistently.
We also improve our ISM model 
by introducing a new way of sampling stars from the stellar initial mass function (IMF) and assigning them to the star particles which allows us to 
form individual massive stars in the simulations.
Finally, 
in addition to the feedback from PE heating and SNe that were already included in Paper I,
we include feedback from photoionization using a Str\"{o}mgren-volume approximation that can deal with overlapping H{\sc ii} regions.
We investigate the impact of stellar feedback on the star formation rate and the galactic outflow rate in dwarf galaxies using different combinations of these feedback processes and examine their systematic differences.
We come to the same conclusion as in Paper I (which is in conflict with \citealt{2016Natur.535..523F})
that supernova feedback plays the dominant role in regulating star formation and shaping the ISM in dwarf galaxies,
and we discuss the potential cause of this conflict in Section \ref{sec:discussion}.

This paper is organized as follows:
In Section \ref{sec:method} we describe the details of our numerical method and our setup of the initial conditions.
In Section \ref{sec:results} we present the results of our numerical simulations.
In Section \ref{sec:discussion} we discuss the implications of our results and the potential caveats.
In Section \ref{sec:summary} we summarize our work.

\section{Numerical Method}\label{sec:method}

\subsection{Gravity and hydrodynamics}

We use the {\sc gadget-3} code \citep{2005MNRAS.364.1105S},
which solves the collisionless gravity with a tree-based method and solves the hydrodynamics with the smoothed particle hydrodynamics (SPH) method \citep{1977AJ.....82.1013L, 1977MNRAS.181..375G}.
We adopt the modified version of SPH described in \citet{2014MNRAS.443.1173H} which has improved accuracy and avoids the numerical artifacts that traditional SPH is prone to develop in multiphase flows. Briefly, we adopt the pressure-energy SPH formulation,
the Wendland $C^4$ kernel with the number of neighboring particles $N_{\rm ngb}$ = 100 in the support radius,
and a switch for the artificial viscosity and artificial thermal conduction devised to only operate at shocks.
Detailed validation tests of this code are presented in \citet{2014MNRAS.443.1173H}.

\subsection{Cooling and chemistry}\label{sec:cooling}

The model of cooling and chemistry we adopt follows Paper I,
which itself is based on  \citet{1997ApJ...482..796N}, \citet{2007ApJS..169..239G} and \citet{2012MNRAS.421..116G}.
We adopt a simplified chemistry network which traces seven chemical species:
H$_2$, H$^+$, CO, H, C$^+$, O and free electrons.
The first three species are explicitly followed while the rest are calculated from the laws for element and charge conservation.
Silicon is assumed to be always in the form of Si$^+$ (singly ionized).
Carbon and oxygen,
if not in the form of CO (carbon monoxide),
are always in the form of C$^+$  and O, respectively.
Our hydrogen chemistry model assumes that H$_2$ can only be formed through the ``dust channel''\footnote{Even at the relatively low metallicities considered in this study, this remains a good approximation; see e.g.\ the discussion in \citet{2003ApJ...584..331G}.},
while it can be destroyed by photodissociation, collisional dissociation and cosmic ray ionization.
The carbon chemistry assumes that CO formation is dictated by the process ${\rm C^+} + {\rm H_2} \rightarrow {\rm CH_2^+}$,
and CH$_2^+$ will either react with O and form CO or be destroyed by photodissociation.

We follow non-equilibrium cooling and heating processes as described in Paper I,
with a major difference that in this work we no longer assume a uniform constant ISRF.
Instead, we calculate the ISRF directly from the stellar particles and therefore make the ISRF spatially and temporally variable (see Section \ref{sec:vG0}).
We apply our non-equilibrium cooling model only if the gas temperature $T$ is lower than $T_{\rm c} =  3\times 10^4$ K.
When $T > T_{\rm c}$, we use to an equilibrium cooling table from \citet{2009MNRAS.393...99W} instead,
which depends on eleven different elements (including hydrogen and helium) that will evolve due to metal enrichment (see Section \ref{sec:SNII}).
In this work we deliberately turn off cosmic ray ionization even though it is included in our ISM model,
in order to focus on the effect of the three feedback mechanisms (described in Section \ref{sec:feedback}) and to aid comparison with \citet{2016Natur.535..523F},

The local dust temperature is calculated following the method of \citet{2012MNRAS.421..116G},
assuming that the dust is in thermal equilibrium where radiative cooling of dust balances dust heating due to absorption of the ISRF, H$_2$ formation on dust and dust-gas collisions.
Under the conditions our simulations explore in this work,
dust heating due to the ISRF is the dominant process.

\subsection{Star formation}

Star formation is a multi-scale process. 
Following the entire dynamical range relevant for star formation is beyond current computational capability, and therefore one always has to resort to sub-resolution models.

On scales below a few parsecs,
the most commonly adopted sub-resolution model is the `sink particle' approach \citep{1995MNRAS.277..362B, 2005ApJ...620L..19L, 2010ApJ...713..269F, 2013MNRAS.430.3261H, 2016arXiv160605346G}.
The idea is that when a region of gas is so gravitationally unstable that it will inevitably collapse and form stars, one can simply convert it into a `sink particle' which only interacts gravitationally but not hydrodynamically with the rest of the system.
Sink particles serve as an efficient device to remove unresolved dense gas from the simulations.
However, 
when the resolution is poor,
not all of the unresolved gas is expected to form stars,
A large fraction of the gas would thus be artificially locked up into sink particles,
if the conventional Jeans criteria for forming sink particles is adopted \citep{1997MNRAS.288.1060B, 1997ApJ...489L.179T}.
Therefore,
the sink particle approach becomes less attractive when the resolution is limited.

On the other hand,
in large-scale cosmological simulations where the resolution is at best $\sim$ 0.1 kpc,
one often adopts a simple star formation recipe based on the Schmidt law:
\begin{equation}
\dot{\rho}_{*} = \epsilon_{\rm sf}  \frac{\rho_{\rm{gas}}}{t_{\rm ff}} \propto \rho_{\rm{gas}}^{1.5},
\end{equation}
where $\dot{\rho}_{*}$ is the instantaneous star formation rate density, $\rho_{\rm{gas}}$ is the gas density, $\epsilon_{\rm sf}$ is an efficiency parameter and $t_{\rm ff}$ is the gas free-fall time \citep{1959ApJ...129..243S, 1992ApJ...391..502K, 1992ApJ...399L.113C}.
Unlike the sink particle approach,
an unresolved gas particle in this recipe would not be locked up into stars instantaneously.
Instead,
it is converted to a star particle with the same mass stochastically on a timescale of $t_{\rm ff} / \epsilon_{\rm sf}$.
Therefore, the unresolved gas has a chance to be dispersed into the diffuse phase again before it forms stars.
Due to the poor resolution in these large-scale simulations ($m_{\rm{gas}} \gtrsim 10^5 M_\odot$),
a star particle usually represents a cluster of stars.
Adopting a simple stellar population model with an assumed IMF,
one can determine the fraction of massive stars and calculate the energy budget of the stellar feedback for each star particle.
For a Kroupa IMF \citep{2001MNRAS.322..231K}, for example,
there is around one type II supernova per 100 M$_\odot$ of stars formed.
A star particle of mass $m_*$ would therefore inject ($m_* / $100 M$_\odot) E_{51}$ of energy into the ISM, where $E_{51} = 10^{51}$ erg is the canonical energy released by a single SN event (see \citealp{2012ARNPS..62..407J}).

However, as the mass of a star particle becomes smaller than 100 M$_\odot$,
the natural minimal unit of SNe is reached and it is unphysical to still inject ($m_* / $100 M$_\odot) E_{51}$ into the ISM for each star particle.
In addition, if ($m_* / $100 M$_\odot) E_{51}$ were injected into the ambient gas with $m_{\rm gas} = m_*$ -- the approach usually adopted in the literature -- then the mass resolution of the hot gas would be too poor to properly resolve the Sedov-Taylor phase of the supernova remnant, leading to serious numerical overcooling. 
This motivates the stochastic feedback injection approach that is adopted in \citet{2012MNRAS.426..140D} and \citet{2016MNRAS.458.3528H},
where each star particle will have a probability of 1/$N_c$ ($N_c > 1$) to inject a feedback energy of $(N_c m_* / $100 M$_\odot) E_{51}$ into the ISM at the end of its lifetime. 
Here $N_c$ is an arbitrary clustering factor which can be chosen as 100 M$_\odot / m_*$ to make sure that each injection represents a SN event.
One can even adopt $N_c >$ 100 M$_\odot / m_*$, which physically means that $N_c m_* / $100 M$_\odot$ SN events are clustered together both spatially and temporally, 
in order to better resolve the Sedov-Taylor phase of the supernova remnant (at the expense of poorer sampling of SN events due to clustering).
However, 
in the stochastic approach the mass injection and the metal enrichment cannot be done consistently with the energy injection, 
simply because a star particle may have less mass and metals than the SN ejecta.

Another caveat of this approach is that it implicitly assumes that the IMF is always uniformly sampled within a star particle,
which is obviously inappropriate when the mass of a star particle becomes smaller than 100 M$_\odot$.
This may not be an issue when the star formation rate is high and a large number of stars are formed locally.
However,
when the local star formation rate is very low,
which is common in dwarf galaxies,
the sampling fluctuations of the IMF can be significant 
(see \citealp{2012ApJ...745..145D, 2014MNRAS.444.3275D} for the effects of the sampling fluctuations).

In this work we adopt the stochastic Schmidt type star formation recipe as described above,
with an efficiency parameter $\epsilon_{\rm sf}$ = 0.02.
However,
instead of interpreting the star particle as a simple stellar population which fairly samples the IMF,
we introduce a flexible method to sample an IMF where each star particle represents an individual star when the resolution is high, 
while in the low resolution limit the simple stellar population model is recovered.
The method is most useful when the mass resolution is close to M$_\odot$,
where the sink particle approach suffers from resolution issues and the simple stellar population does not apply.


\subsubsection{Sampling the stellar masses from an IMF}\label{sec:IMFsample}
In this model,
a gas particle is first converted to a star particle of the same mass using the stochastic star formation recipe.
Then,
for each star particle of mass $m_*$ ($m_* = m_{\rm gas})$, we randomly draw an array of stellar masses ($m_{\rm IMF}$) from the assumed IMF and assign them to the star particle until the total sampled stellar mass $M_{\rm IMF} = \sum m_{\rm IMF}$ equals or exceeds $m_*$.
In practice, $M_{\rm IMF}$ is unlikely to exactly equal $m_*$, and there will always be a residual mass $M_{\rm IMF} - m_*$, 
which we transfer to the next star particle,
so that the total mass conservation is ensured.\footnote{Note that we cannot simply require that $M_{\rm IMF} = m_{*}$ as an additional constraint during the sampling, as this leads to us over-sampling low mass stars and under-sampling high mass stars \citep{2016arXiv161002538S}.}
After the sampling is done,
we modify the mass of the star particle to be $M_{\rm IMF}$.
When the star particles are massive (poor resolution),
the residual mass will only be a small fraction of the original $m_*$,
so the star particles will still have almost equal mass after the sampling.
On the other hand, 
when the star particles have masses of the order of a solar mass,
even one massive O star can greatly exceed $m_*$.
In this case,
the subsequent star particles will be assigned with zero mass until the residual mass is exhausted.
Star particles that are assigned with zero mass will be removed from the simulations.
We note that modifying the mass implies an unphysical mass transfer between star particles,
so some sort of distance constraint for the mass transfer is desirable,
though it is not yet implemented in this work.

We adopt the Kroupa IMF and set the maximum and minimum masses to be 50 M$_\odot$ and 0.08 M$_\odot$, respectively.
To sample random numbers following the Kroupa IMF, we use the acceptance-rejection method with a power-law envelope function, which has a high acceptance rate of about 65\%.
For the Kroupa IMF and our adopted mass range,
the average number of stars per solar mass is 1.819,
which means that the average number of stars assigned in each star particle would be 1.819 $m_* / {\rm M}_\odot$ with certain fluctuations.
However,
since stars below 1 M$_\odot$ do not contribute much to both the UV luminosity and the metal enrichment,
we do not store them into the stellar mass array in order to save computer memory 
(but we still add them to $M_{\rm IMF}$).
The average number of stars above 1 M$_\odot$ is 0.175 $m_* / {\rm M}_\odot$, which is about ten times less memory demanding.

\subsubsection{The star formation threshold}\label{sec:SF_th}

Unlike in Paper I, where we adopted both density and temperature thresholds for star formation,
in this work we adopt a Jeans star formation threshold which only involves one free parameter.
For every gas particle, its associated Jeans mass is defined as
\begin{equation}
	M_{{\rm J},i} = \frac{\pi^{5/2} c_{s,i}^3}{6 G^{3/2} \rho_i^{1/2}}
\end{equation}
where $G$ is the gravitational constant, and $c_{s,i}$ and $\rho_i$ are the sound speed and the density of the gas particle $i$, respectively.
A gas particle $i$ will be assigned with a non-zero instantaneous star formation rate and will be called ``star-forming'' only if $M_{{\rm J},i} < N_{\rm th} M_{\rm ker}$,
where $M_{\rm ker} = N_{\rm ngb} m_{\rm gas}$ is the SPH kernel mass and $N_{\rm th}$ is a free parameter.
In this work we choose $N_{\rm th}$ = 8 in order to properly resolve the Jeans mass for the star-forming gas.
The effect of varying $N_{\rm th}$  is investigated in Appendix \ref{app:MJtest}.
In addition,
we further require that the gas has to have a negative local velocity divergence (i.e., converging flows) in order to be star-forming.

\subsection{Interstellar UV radiation field}\label{sec:vG0}

We adopt an approximation similar to \citet{2016Natur.535..523F} which is only valid in dust-poor systems.
In galaxies with a dust-to-gas ratio (DGR) close to that of the Milky Way,
calculating a local ISRF is nontrivial because of dust extinction, 
and one usually has to resort to radiative transfer which is very computationally expansive.
However,
in dwarf galaxies where the DGR is significantly lower,
dust extinction becomes negligible and one can simply calculate the local ISRF by summing up all contributions from the stellar sources using the inverse square law.

For a given sampled stellar mass, 
we obtain the effective temperature and the stellar lifetime using the results of \citet{2013A&A...558A.103G},
which assume a metallicity $Z$ = 0.002 ($\sim 0.1 Z_\odot$).
We then adopt the stellar spectra from the BaSeL stellar library \citep{1997A&AS..125..229L, 1998A&AS..130...65L, 2002A&A...381..524W}
and integrate the spectra to obtain the UV luminosity in the energy range of 6 $-$ 13.6 eV,
which is relevant for the photoelectric heating in the ISM.
The total UV luminosity of a star particle is then obtained by summing over the contributions from all the assigned stellar masses whose ages are younger than their stellar lifetimes.

The energy density at the location of a gas particle is calculated by
\begin{equation}\label{eq:u_tree}
	u_{6-13.6 {\rm eV}} = \sum_{i} \frac{L_{i, 6-13.6 {\rm eV}}}{4\pi c r_i^2},
\end{equation}
where $c$ is the speed of light, $L_{i, 6-13.6 {\rm eV}}$ is the UV luminosity of star particle $i$, and $r_i$ is the distance between the gas particle and the star particle 
\footnote{A star particle $i$ may contain several stars (see Section \ref{sec:IMFsample}) and we sum over their contributions to obtain $L_{i, 6-13.6 {\rm eV}}$. Physically, $L_{i, 6-13.6 {\rm eV}}$ mainly comes from stars with $m_{\rm IMF} > 4$ M$_\odot$ which traces the recent star formation on the timescale of 100 Myr, though in practice we include all stars with $m_{\rm IMF} > 1$ M$_\odot$.}.
We take the advantage of the gravity tree and store the UV luminosity into tree nodes.
This allows us to perform the summation in Eq. \ref{eq:u_tree} while walking the tree, which greatly reduces the computational cost.
The tree approach gives exact results in the vicinity of the star and starts to show some small deviation (controlled by the opening criteria of the gravity tree) at larger distances due to the tree approximation,
which is sufficient for our purpose.

\subsection{Stellar feedback}\label{sec:feedback}

\subsubsection{Photoelectric heating}


As in Paper I,
we adopt the following PE heating rate \citep{1994ApJ...427..822B, 2003ApJ...587..278W, 2004ApJ...612..921B}:
\begin{equation}\label{eq:photoheat}
\Gamma_{\rm pe} = 1.3\times 10^{-24} \epsilon D G_{\rm eff}  n  ~{\rm erg~ s^{-1} cm^{-3}},
\end{equation}
where $G_{\rm eff} = G_0 {\rm exp}(-1.33\times 10^{-21} D N_{\rm H,tot})$ is the attenuated radiation field strength in units of the \citet{1968BAN....19..421H} field, $n$ is the hydrogen nuclei number density and $\epsilon$ is the photoelectric heating efficiency defined as
\begin{equation}
\epsilon = \frac{0.049}{1 + (0.004\psi^{0.73})} 
+ \frac{0.037 (T/10000)^{0.7}}{1 + 2\times 10^{-4} \psi}
\end{equation}
where $\psi = G_{\rm eff} T^{0.5}/n_{\rm e^-}$ and $n_{\rm e^-}$ is the electron number density.
The major difference is that in this work $G_0$ is locally obtained by 
\begin{equation}
	G_0 = \frac{u_{6-13.6 {\rm eV}} }{5.29\times 10^{-14}\ {\rm erg\ cm^{-3}}},
\end{equation}
where $u_{6-13.6 {\rm eV}} $ is calculated via Eq. \ref{eq:u_tree}
which varies both spatially and temporally.
In the solar neighbourhood,
the estimated $G_0$ is 1.7 \citep{1978ApJS...36..595D}. 
We set a minimum value of $G_0 \geq G_{\rm 0,min} = $ 0.00324, which is the cosmic UV background from \citet{2012ApJ...746..125H} integrated over the 6--13.6~eV energy range.

\begin{figure}
	\centering
	\includegraphics[width=1\linewidth]{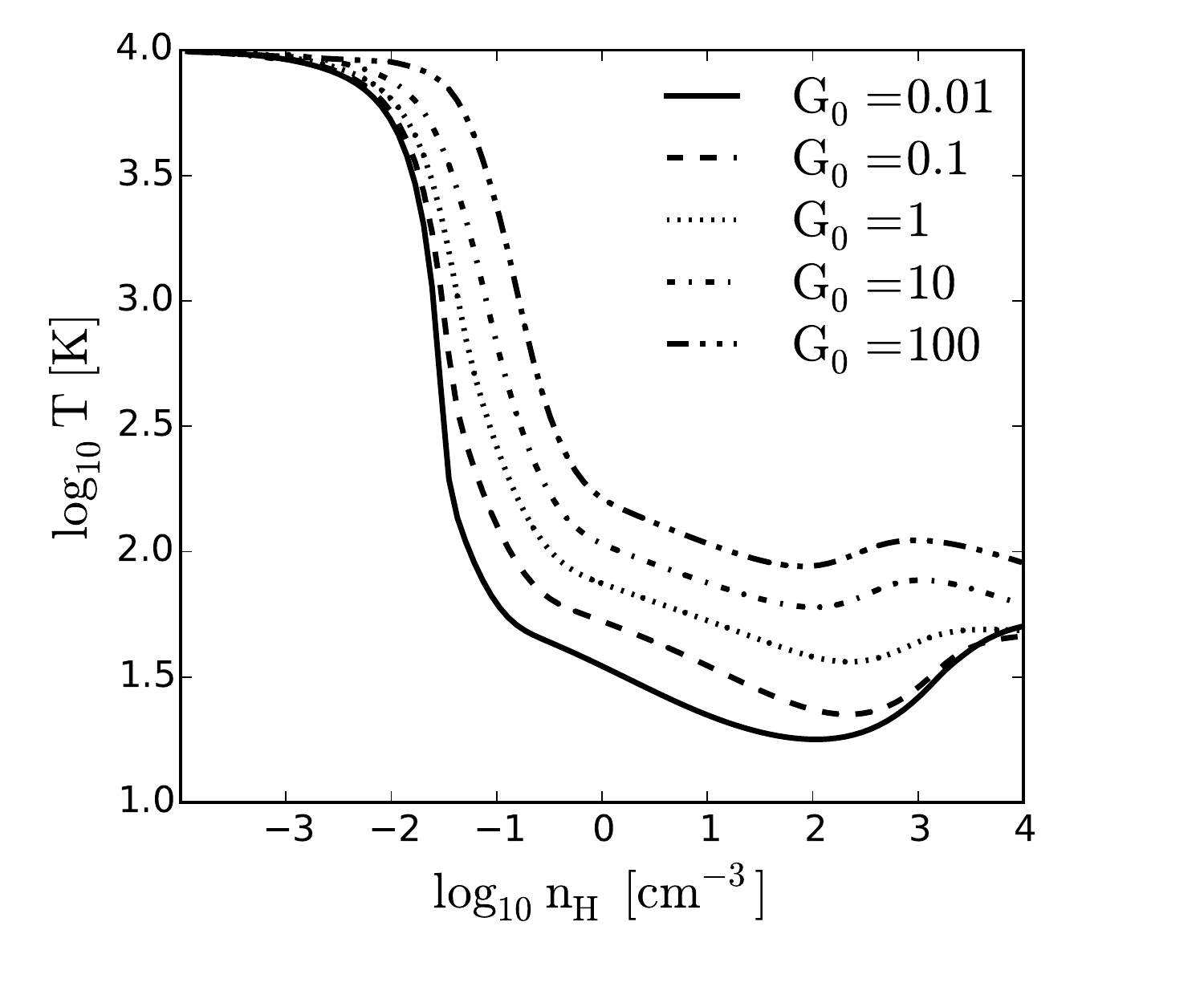}
	\caption{
		Temperature vs. hydrogen number density for gas in thermal equilibrium under five different values of $G_0$,
		The adopted metallicity is 0.1 $Z_\odot$ and the dust-to-gas mass ratio is 0.1 \%.
		Heating from cosmic ray ionization is deliberately switched off for simplicity and to aid comparison with \citet{2016Natur.535..523F}.
		Increasing $G_0$ leads to a higher equilibrium temperature due to more efficient PE heating.
		In contrast to photoionization, which can heat dense gas up to $\sim 10^4$ K,
		PE heating only heats dense gas up to a few hundred Kelvin even in the most extreme case.
	}
	\label{fig:PD_eq_5}
\end{figure}

Fig. \ref{fig:PD_eq_5} shows the gas (thermal-)equilibrium temperature ($T$) as a function of its hydrogen number density ($n_{\rm H}$) for several different strengths of the ISRF, assuming that the gas is optically thin. Increasing the ISRF leads to a larger PE heating rate and therefore a higher equilibrium temperature.
As mentioned in Section \ref{sec:cooling}, the cosmic ray ionization heating is not included throughout this work for simplicity and to aid comparison with \citet{2016Natur.535..523F}.
In the range of $n_{\rm H} = 0.01 - 0.1$ cm$^{-3}$,
the equilibrium curves have a steep negative slope where $T$ drops sharply as $n_{\rm H}$ increases.
In the ``cold phase'' regime ($n_{\rm H} \gtrsim 1 {\rm cm^{-3}}$),
the scaling between $G_0$ and $T$ is sub-linear: increasing $G_0$ by four orders of magnitude only leads to less than one order of magnitude increase in equilibrium temperature.

The treatment of radiation shielding is the same as  Paper I,
adopting the {\sc TreeCol} algorithm \citep{2012MNRAS.420..745C} to estimate the column densities along twelve different lines of sight for each gas particle.
The column densities are accumulated by taking advantage of the gravity tree structure up to a predefined shielding length $L_{\rm sh}$ = 50 pc.

\subsubsection{Photoionization}\label{sec:PI}

Massive stars emit UV radiation that ionizes and heats up the ambient medium. 
This can lead to gas expansion in the immediate vicinity of the stars.
As a result,
when they explode as supernovae, they do so in a low density environment, rather than a dense
molecular cloud. It is therefore important to account for photoionization feedback if we want to properly capture the effects of
the supernovae \citep[see e.g.][]{2016arXiv161006569P}. 

Here we adopt a Str\"{o}mgren type approximation for the photoionization similar to \citet{2012MNRAS.421.3488H}.
For a given star particle with at least one assigned stellar mass $m_{\rm IMF} >$ 8 M$_{\odot}$,
we flag its neighboring gas particles within a certain radius $R_{\rm s}$ as ``photoionized''
and set their ionization fraction to be 1. 
These gas particles are not allowed to cool below a minimum temperature of $10^4$ K,
the typical temperature in an HII region.

In a medium of uniform density,
$R_{\rm s}$ can be determined using the classical Str\"{o}mgren radius $R_{\rm s} = (3 S_*/ 4\pi n^2 \beta)^{1/3}$, 
where $S_{*}$ is the ionizing photon rate of the star,
$\beta$ = 2.56 $\times 10^{-13}$ cm$^3$ sec$^{-1}$ is the case B hydrogen recombination coefficient, evaluated at $T = 10^{4}$~K,
and $n = \rho_{\rm g} X_{\rm_H} / m_{\rm p}$ is the hydrogen nuclei number density where $X_{\rm_H}$ is the hydrogen mass faction and $m_{\rm p}$ is the proton mass.
However,
if the density is inhomogeneous,
the gas density estimated at the location of the star may be very different from the density of the neighboring gas relevant for recombination.
Moreover,
when two Str\"{o}mgren spheres overlap (e.g. two stars at the same location),
this simple method would double-count the photoionized gas and therefore underestimate $R_{\rm s}$.
As such,
we determine $R_{\rm s}$ iteratively as follows.

Assuming ionization equilibrium,
the hydrogen recombination of the neighboring gas should balance the ionizing photon budget of the star.
Each neighboring gas particle contributes $\beta n_{\rm H} N_{\rm H}$ recombinations per second,
where $N_{\rm H} = m_{\rm g} X_{\rm_H} / m_{\rm p}$ is the number of hydrogen nuclei for the gas particle.
For each star particle, we first search for a few ($\sim$ 10) neighboring gas particles and calculate the total recombination rate $R_{\rm rec} = \beta \sum_i (n^i_{{\rm H}} N^i_{\rm H})$, where the summation is over the neighboring particles $i$.
To cope with overlapping HII regions,
if a neighboring gas particle is already flagged as ``photoionized'' by another star particle,
it will be excluded from the summation.
We note that this process depends on the order of flagging for different star particles which may affect the reproducibility of the results \footnote{Our neighbor search is done globally instead of only on the local domain, which prevents artificial cut-off by domain decomposition.},
though statistically this is not a major concern.
If $S_{*} - R_{\rm rec} > 0$,
we look for more neighbors by increasing the searching radius by 10 \% and repeat the flagging,
which is analogous to propagating the ionization front.
If $S_{*} - R_{\rm rec} < 0$,
we decrease the searching radius by 10 \% and unflag the particles that are now outside of the radius.
We do so iteratively until $|S_{*} - R_{\rm rec}|$ is smaller than some tolerance, which is set to be $10 \beta m_{\rm g, in}X_{\rm_H} / m_{\rm p}$.
This tolerance is chosen to be resolution dependent and it corresponds to the recombination rate of one gas particle in an ambient hydrogen density of 10 cm$^{-3}$.
In addition,
if the current iteration only flags/unflags one more gas particle compared to the previous iteration and $S_{*} - R_{\rm rec}$ changes sign,
we terminate the iteration as it means that this gas particle has a density that is too high such that the stopping criterion will never be met: performing more iterations would only repeatedly flag and unflag this gas particle.
In Appendix \ref{app:Spitzer},
we show that this approach converges and agrees well with that predicted by more sophisticated radiative transfer methods in a uniform medium,
though in an inhomogeneous medium the method is expected to be strongly mass-biased.

Although our approach is more accurate than using the simple Str\"{o}mgren radius as described above,
it is still strongly mass-biased, i.e., it preferentially ionizes mass concentrated regions.
This is because our approach ignores any angular information. For example,
  consider the case of a dense molecular cloud located at some distance from an ionizing
  source that is otherwise surrounded only by diffuse gas. The Lagrangian nature
  of SPH means that in this situation, the majority of the SPH particles in the 
  vicinity of the ionizing source may be contained within the dense cloud. 
  In our method, the cloud would therefore be assigned the majority of the ionizing flux, 
  even though from the point of view of the ionizing source it may subtend only a
  small solid angle. One potential improvement in a future version of this method 
  would be to include at least some angular information by subdividing the $4\pi$ solid 
  angle around each ionizing source into several pixels and applying the method
  separately to each pixel. For the time being, however, we prevent the method from
  artificially ionizing a dense cloud far away from the star by imposing a maximum $R_{\rm s}$ of 50 pc.
This (purely empirical) value is motivated by the typical size of HII regions deduced from observations of nearby dwarf galaxies (see e.g. \citealp{2015A&A...578A..53C}).
Despite the approximate nature of this method,
it captures the physical behavior of photoionization at a low computational cost compared to the radiative transfer approach
(see e.g. \citealp{2015MNRAS.451...34R} and \citealp{2016arXiv161006569P}).

\subsubsection{Supernovae and metal enrichment}\label{sec:SNII}

A star particle with at least one assigned stellar mass $m_{\rm IMF} >$ 8 M$_{\odot}$ will explode as a supernova type II (SNII) at the end of its lifetime obtained from \citet{2013A&A...558A.103G}.
We further restrict the timesteps of massive star particles to a maximum of 0.1 Myr, in order to make sure that they explode promptly after the end of their lifetimes.
In practice, however, the typical timesteps of stars are usually smaller than 0.1 Myr due to other timestep constraints (e.g. the gravitational acceleration).
Regardless of the actual stellar mass, each SNII injects $10^{51}$ erg of thermal energy into its nearest 100 particles weighted by a cubic spline kernel function\footnote{For the energy injection, the adopted kernel function has little effect on the results as opposed to the SPH interactions where the cubic spline kernel function suffers from the numerical clumping instability.
}.
We do not include type Ia supernovae in this work.

We adopt the metal yields of SNII from \citet{2004ApJ...608..405C} to account for the metal enrichment.
The mass of the total ejecta and individual elements are added to the nearest 100 gas particles, also weighted by the cubic spline kernel.
Since the average mass of the gas particle can be much smaller than the ejecta mass of massive stars,
the neighboring gas particles can be much more massive than the rest of the gas particles after the enrichment,
which will lead to undesirable numerical noise in SPH.
Therefore,
we split a gas particle into two particles whenever its mass becomes larger than $2 m_{\rm gas,init}$ where $m_{\rm gas,init}$ is the gas particle mass in the initial conditions,
and we do so iteratively until all gas particles have their masses smaller than $2 m_{\rm gas, init}$ to make sure that this is the maximum mass of gas particles at all times.
The two split particles inherit all physical quantities from their parent particle 
(including the velocity, specific thermal energy, metallicity and chemical abundances) 
except for their mass, which is half the parent particle mass, and their positions, 
which are offset symmetrically by one-fifth of the smoothing length of the parent particle in a random direction.

\begin{table*}
	\caption{Simulation runs and the corresponding setups. 
		Feedback processes that are included in the simulation runs are marked as `yes', and `no' otherwise.
		$l_{\rm gas}$ and $l_{\rm star}$ are the scale-lengths of the gaseous disk and the preexisting stellar disk, respectively, where
		$l_{\rm star}$ = 0.73 kpc in all runs.
		\textit{cPE-PI-SN} has a constant radiation field with $G_0$ = 0.1 instead of a spatially and temporally varying ISRF.
		\textit{PE-PI-SN-3Myr} has a constant stellar life time of 3 Myr for all massive stars.
	}  
	\label{table:parameters}
	\begin{tabular}{| l | l | l | l | l | l | l |}    
		\hline\hline
		Name                   & photoelectric heating & photoionization & supernovae & $l_{\rm gas}/l_{\rm star}$ & Notes &\\
		\hline
		\textit{PE-PI-SN}            & yes & yes & yes & 2\\
		\textit{PE-noPI-SN}       & yes & no  & yes & 2\\
		\textit{noPE-PI-SN}       & no  & yes & yes & 2\\
		\textit{noPE-noPI-SN}       & no  & no  & yes & 2\\
		\textit{PE-noPI-noSN}   & yes & no  & no  & 2\\
		\textit{cPE-PI-SN}          & yes & yes  & yes & 2 & $G_0$ = 0.1 (constant)\\
		\textit{PE-PI-SN-3Myr} & yes & yes & yes & 2 & $t_{\rm life}$ = 3 Myr (constant)\\
		\hline
		\textit{PE-PI-SN-cmp}           & yes & yes & yes & 1\\
		\textit{noPE-PI-SN-cmp}       & no  & yes & yes & 1\\				
		\textit{PE-noPI-noSN-cmp}  & yes & no  & no  & 1\\				
		\textit{noPE-noPI-SN-cmp}  & no & no  & yes  & 1\\				
		\hline\hline
	\end{tabular}
\end{table*}

\subsection{Initial conditions}\label{sec:ICs}
We set up the initial conditions using the code developed by \citet{2005MNRAS.361..776S}.
The dark matter halo has a virial radius $R_{\rm vir}$ = 44 kpc and a virial mass $M_{\rm vir}$ = $2\times 10^{10} {\rm M}_\odot$.
It follows a Hernquist profile with an NFW-equivalent \citep{1997ApJ...490..493N} concentration parameter $c$ = 10 
and the spin parameter $\lambda$ = 0.03.
A preexisting stellar disk of $2\times 10^7 {\rm M}_\odot$ and a gas disk of $4\times 10^7 {\rm M}_\odot$ are embedded in the dark matter halo,
which results in a low baryonic mass fraction of 0.3 \% that is motivated by abundance matching \citep{2010ApJ...710..903M, 2013MNRAS.428.3121M}.
The stellar disk follows an exponential profile with scale-length $l_{\rm star}$ = 0.73 kpc and scale-height of 0.35 kpc.
The gas disk also follows an exponential profile,
and in this work we explore two different initial conditions that differ only in their gas disk scale-lengths $l_{\rm gas}$: one with 0.73 kpc and the other with 1.46 kpc.
The latter will be our fiducial choice as the gas disk is usually observed to be more extended than the stellar disk \citep{2008MNRAS.386.1667B,
 2012AJ....144..134H}.
The difference in $l_{\rm gas}$ leads to a factor of $3$--$4$ difference in gas surface density in the central region of the disk (see Fig. \ref{fig:IC_profile}).
The scale-height of the gas disk is determined assuming vertical hydrostatic equilibrium.

The initial gas temperature is set to be $10^4$ K.
The initial metallicity (both gas and stars) is set to be 0.1 Z$_\odot$ uniformly, 
and the relative abundances of the eleven elements follow the solar abundances.
The dust-to-gas mass ratio is set to be 0.1 \%.
We adopt a particle mass of $m_{\rm dm} = 10^4\ {\rm M}_\odot$ for the dark matter, $m_{\rm disk} = 4\ {\rm M}_\odot$ for the stellar disk, and $m_{\rm gas} = 4\ {\rm M}_\odot$ for the gas.  The gravitational softening length is 62 pc for the dark matter and 2 pc for the baryons.

We use cylindrical coordinates $R$ and $z$ to describe the simulations,
where $R$ is the galactocentric radius and $z$ is along the rotation axis of the disk.
The origin is chosen at the center of mass of the stellar disk.
We define the ``ISM region'' as  $R <$ 1.5 kpc and $|z| < $ 1 kpc where the star formation takes place.
The simulation time is denoted as $t$.

\subsubsection{Initial turbulent driving}\label{sec:driving}

\begin{figure}
	\centering
	\includegraphics[trim = 10mm 0mm 25mm 0mm, clip, width=0.9\linewidth]{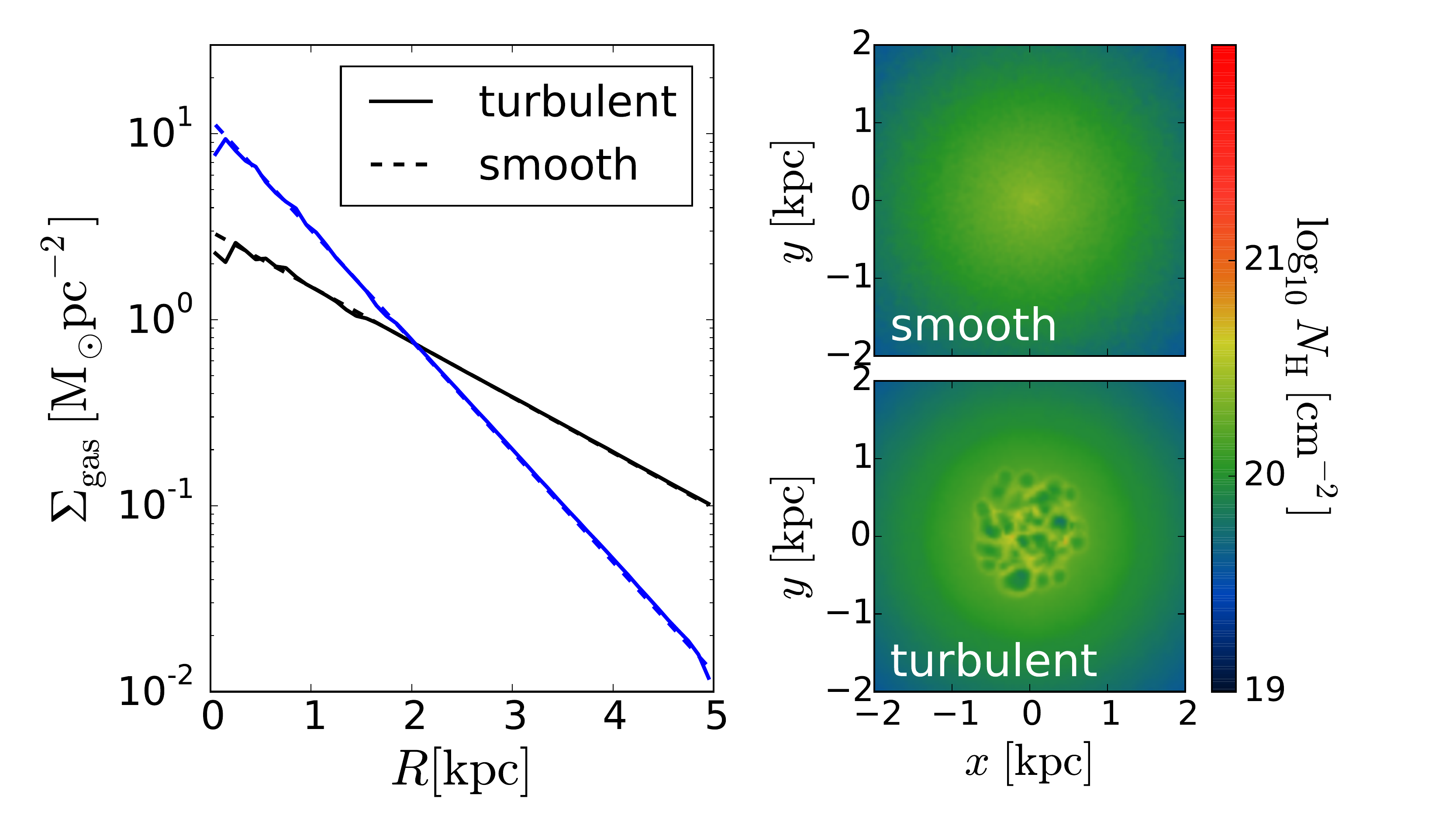}
	\caption{
		Left panel:
		gas surface density ($\Sigma_{\rm gas}$) vs. galactocentric radius ($R$) in the initial conditions, with a binsize of 0.1 kpc. 
		Our fiducial setup (black) has  a scale-length of the gas disk $l_{\rm gas}$ = 1.46 kpc, while the alternate setup (blue) which is referred to as the `compact disk' has $l_{\rm gas}$ = 0.73 kpc.
		The gas surface density in the fiducial setup is lower than in the alternate setup in the entire region of $R<$ 1.5kpc where star formation occurs.
		The dashed lines show the smooth exponential radial profiles while the solid lines are the profiles after the initial turbulence is introduced, which only slightly affects the large-scale profile.
		Right panels:
		column density maps for the fiducial disk before (top right) and after (bottom right) the preexisting turbulence is introduced,
		which seeds the small-scale density perturbations but keeps the large-scale profile unchanged.		
		}
	\label{fig:IC_profile}
\end{figure}

With the smooth initial conditions described above, we have seen that the gas quickly collapses along the $z$-axis to form a very thin layer, 
which in turn leads to a subsequent burst of star formation (for around 0.1 Gyr or so) that creates a hole in the central part of the galaxy (see Paper I).
We consider this a numerical artifact which motivated us to introduce preexisting turbulence in our gas disk by running a setup simulation for 20 Myr.
In the setup simulation we switch off star formation and stellar feedback,
but we inject thermal energy into the gas disk in a way that mimics the effect of SNe:
we first identify locations of overdense gas particles (defined as $n_{\rm gas} >$ 0.1 cm$^{-3}$ and $T > 10^4$ K) within the central region $R < 1$ kpc
and then stochastically inject $10^{51}$ erg at each of these locations and its associated 100 neighboring particles.
The probability to inject this thermal energy for each overdense gas particle within a timestep of $\Delta t$ is $0.02 \Delta t / t_{\rm ff}$,
in the same spirit as our stochastic star formation model.
By doing so we obtain a turbulent gas disk with preexisting structures as shown in Fig. \ref{fig:IC_profile}.
We found it to be an efficient way to avoid the initial burst of star formation and outflows, allowing the system to settle into a steady state more rapidly.

\subsubsection{Naming conventions}\label{sec:naming}
We run ten simulations with different combinations of stellar feedback, stellar lifetimes and initial conditions.
The naming convention is listed in Table \ref{table:parameters},
where \textit{PE}, \textit{PI} and \textit{SN}  stands for photoelectric heating, photoionization and supernovae (type II), respectively.
The PE heating is switched off by setting $G_0$ = 0 (instead of $G_0 = G_{\rm 0,min}$) everywhere.
We also run a simulation with a constant $G_0$ = 0.1 (denoted as \textit{cPE}). 
The value is chosen to be linearly scaled with the star formation rate surface density, which in our fiducial simulation is roughly an order of magnitude lower than in the
  local ISM.\footnote{Recall that $G_{0} \sim 1$ in the solar neighborhood.}  Finally, we run a simulation where all the massive stars ($m_{\rm IMF} > 8 {\rm M}_\odot$) have a constant stellar lifetime of 3 Myr (denoted as \textit{3Myr}).
The initial condition with $l_{\rm gas}$ = 0.73 kpc will be referred to as the ``compact disk'' and denoted as \textit{cmp} 
while the one with $l_{\rm gas}$ = 1.46 kpc will be referred to as the ``fiducial disk''.

\begin{figure*}
	\centering
	\includegraphics[trim = 10mm 20mm 0mm 0mm, clip, width=1.0\linewidth]{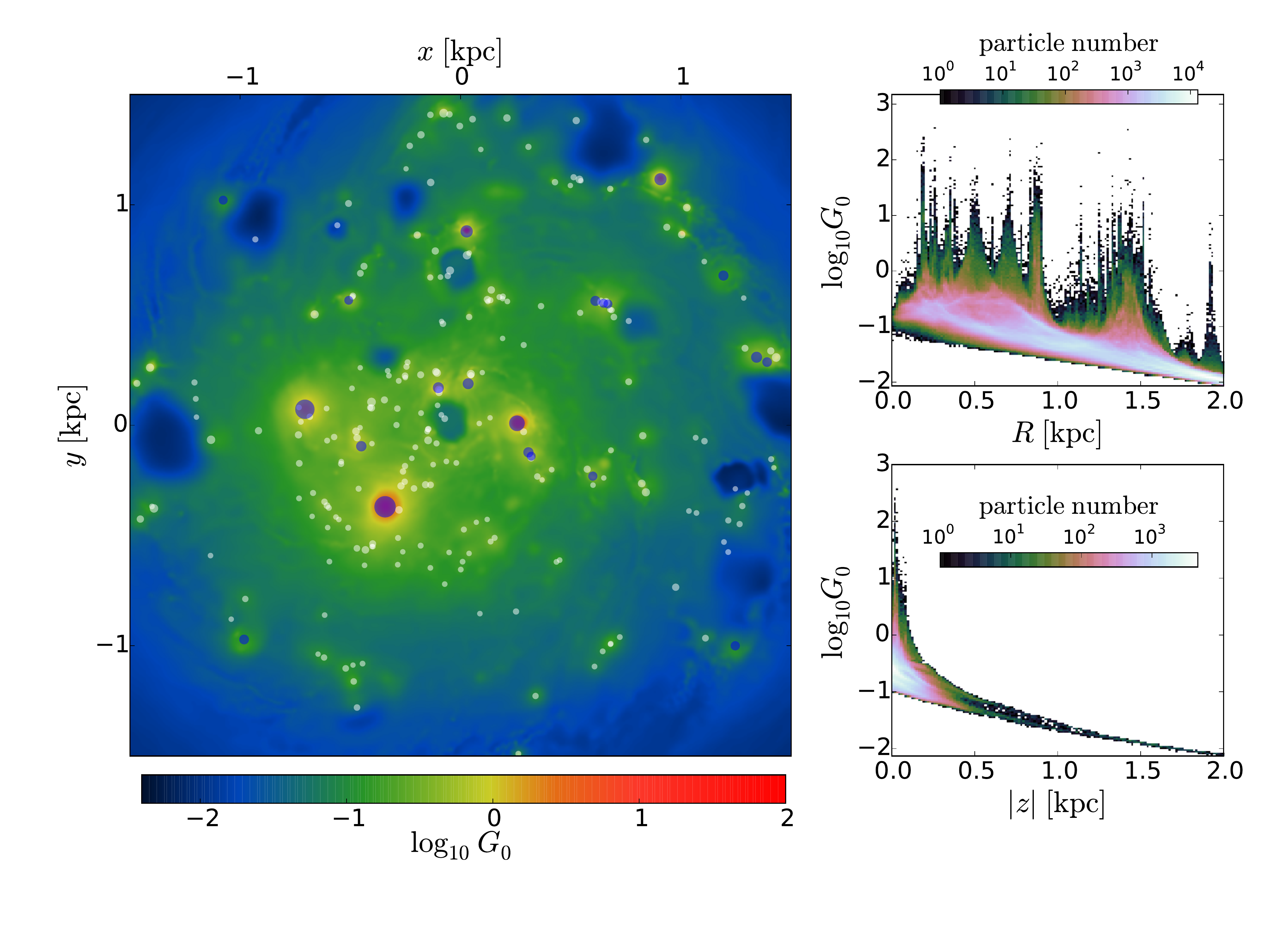}
	\caption{
		Left panel: a face-on map of the (mass-weighted) ISRF at $z$ = 0 for the fiducial run \textit{PE-PI-SN} at simulation time $t$ = 600 Myr. 
		The star particles with $m_* > 8{\rm M}_\odot$ and $4 {\rm M}_\odot < m_* \leq 8 {\rm M}_\odot $ are overplotted in blue and white circles, respectively.
		Their sizes are linearly proportional to their masses.
		Note that the holes that associate with the feedback-driven bubbles in the map are mostly visualization artifacts as there are very few particles within the holes.
		Top right panel: the ISRF as a function of $R$ for gas within $|z| <$ 0.5 kpc.
		Bottom right panel: the ISRF as a function of $|z|$ for gas within $R <$ 0.5 kpc.}
	\label{fig:g0mapprofile}
\end{figure*}

\section{Results}\label{sec:results}

\subsection{The interstellar radiation field}

The left panel of Fig. \ref{fig:g0mapprofile} shows a face-on map of the (mass-weighted) ISRF as seen by the gas at $z$ = 0 for the fiducial run \textit{PE-PI-SN} at $t$ = 600 Myr.
The star particles with $m_* > 8{\rm M}_\odot$ and $4 {\rm M}_\odot < m_* \leq 8 {\rm M}_\odot $ are overplotted with blue and white circles, respectively.
Their sizes are linearly proportional to their masses.
Stars less massive than $4 {\rm M}_\odot$ do not contribute much to the ISRF and hence are not plotted.
Note that since we adopt $m_{\rm gas} = 4 {\rm M}_\odot $ these star particles represent individual stars.

A smooth background radiation field that gradually declines with $R$ is naturally established from the UV-emitting star particles.
In addition to the smooth background,
the ISRF can be locally enhanced significantly in the vicinity of UV-emitting stars.
This is especially visible  for massive stars ($m_* > 8{\rm M}_\odot$) which have higher luminosities,
though they are very few in number both because of the IMF and because they are short-lived.
Note that since the map is mass-weighted (hence mass-biased) the ``holes'' that associate with the feedback-driven bubbles in the map are mostly visualization artifacts.
To be more quantitative,
the top right panel of Fig. \ref{fig:g0mapprofile} shows the ISRF as a function of $R$ for gas within $|z| <$ 0.5 kpc,
while the bottom right panel shows the ISRF as a function of $|z|$ for gas within $R <$ 0.5 kpc.
Here the ISRF is shown directly on the particle level without mass weighting,
and therefore it shows no artificial deficit within the holes.
The majority of the gas is subject to the smooth ISRF,
which declines with $R$ from $G_0 \sim 0.2$ in the central region to $G_0 \sim 0.01$ at $R$ = 2 kpc.
Along the z-axis the ISRF declines very rapidly as most star formation occurs within the mid-plane.
Only a small number of gas particles close to the UV-emitting stars have locally enhanced ISRF which can be orders of magnitude higher than the background field.

\begin{figure*}
	\centering
	\includegraphics[trim = 30mm 0mm 30mm 0mm, clip, width=1.0\linewidth]{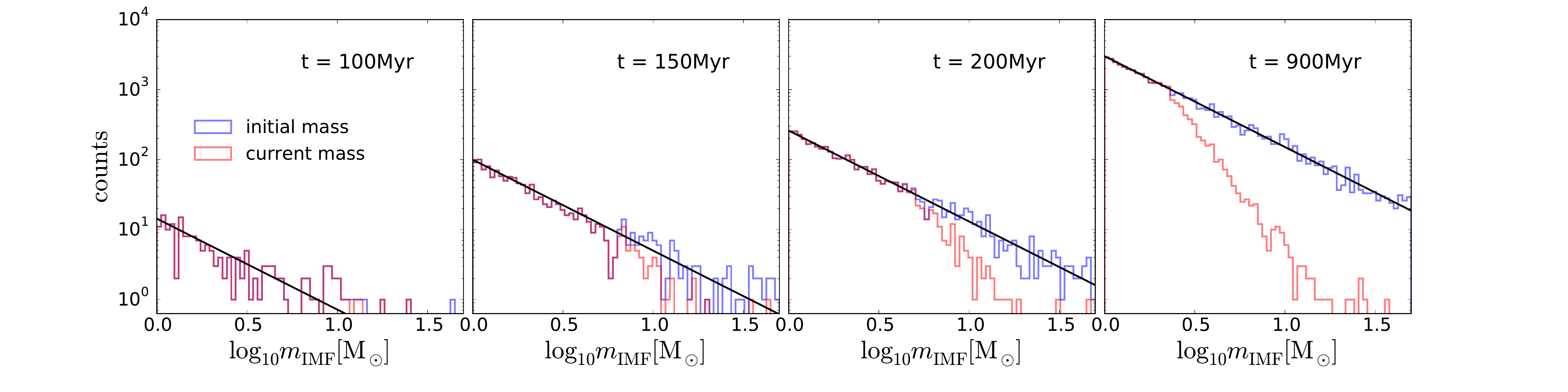}
	\caption{
		Histograms of stellar masses ($m_{\rm IMF}$) drawn from the Kroupa IMF at four different simulation times in the \textit{PE-PI-SN} run.
		Note that multiple stellar masses can be stored in one star particle until $\sum m_{\rm IMF} \geq m_*$.
		The initial masses (including stars that have already evolved past their lifetimes) are shown in blue, 
		while stars that are still alive at time $t$ are shown in red. 
		The statistical fluctuation from the power-law IMF decreases as time passes and more stars form.}
	\label{fig:m_imf_time}
\end{figure*}

\subsubsection{Stellar evolution}

Fig. \ref{fig:m_imf_time} shows the histograms of stellar masses ($m_{\rm IMF}$) drawn from the Kroupa IMF at four different simulation times in the  \textit{PE-PI-SN} run.
Note that multiple stellar masses can be stored in one star particle until $\sum m_{\rm IMF} \geq m_*$.
The initial masses (including stars that have already evolved past their lifetimes) are shown in blue, which show large deviations from the power-law IMF at $t$ = 100 Myr due to statistical fluctuation.
As the total number of stars increases with time the fluctuation decreases and the IMF becomes well-sampled.
The red histograms represent the stars that are still alive at time $t$.
Since the stellar lifetime is a monotonically decreasing function of the stellar mass,
the red histograms deviate from the blue ones at decreasing stellar mass as time evolves.
However,
due to the continuous star formation,
there are still massive stars that formed recently instead of a sharp cut-off in the histogram.

\begin{figure*}
	\centering
	\includegraphics[trim = 30mm 10mm 40mm 20mm, clip, width=0.95\linewidth]{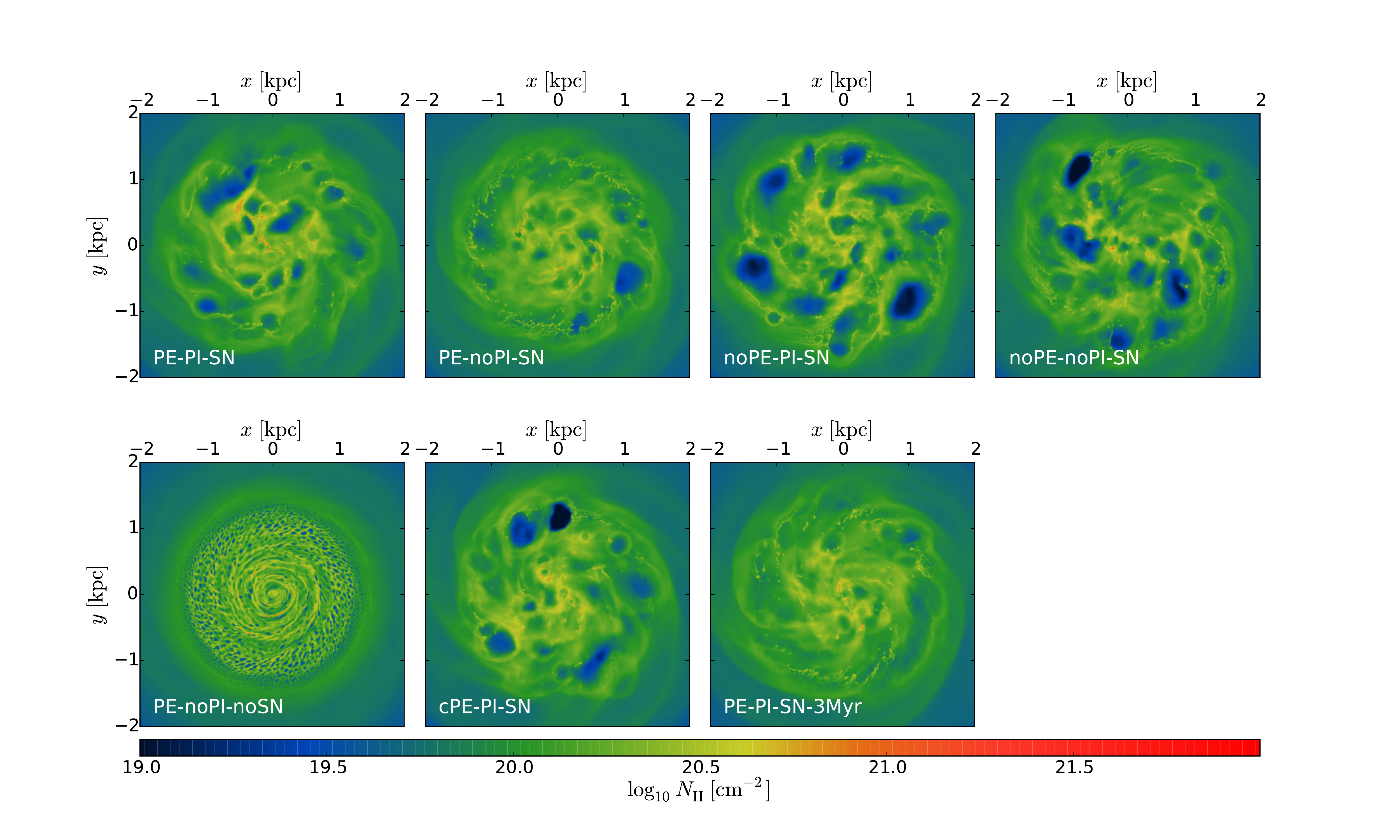}
	\caption{
		Face-on maps of the hydrogen nuclei column density of all runs for the fiducial disk at $t$ = 500 Myr.
		The PE-only run (\textit{PE-noPI-noSN}) shows a drastically different morphology from the other runs. 
		The feedback-driven bubbles show different sizes in different runs.
		In particular, there are almost no bubbles in \textit{PE-PI-SN-3Myr}.
		As will be discussed in Section \ref{sec:outflows}, 
		this is because very few SNe go off at low densities in this run.}
	\label{fig:faceOnMaps500Myr}
\end{figure*}

\begin{figure*}
	\centering
	\includegraphics[trim = 30mm 10mm 40mm 20mm, clip, width=0.95\linewidth]{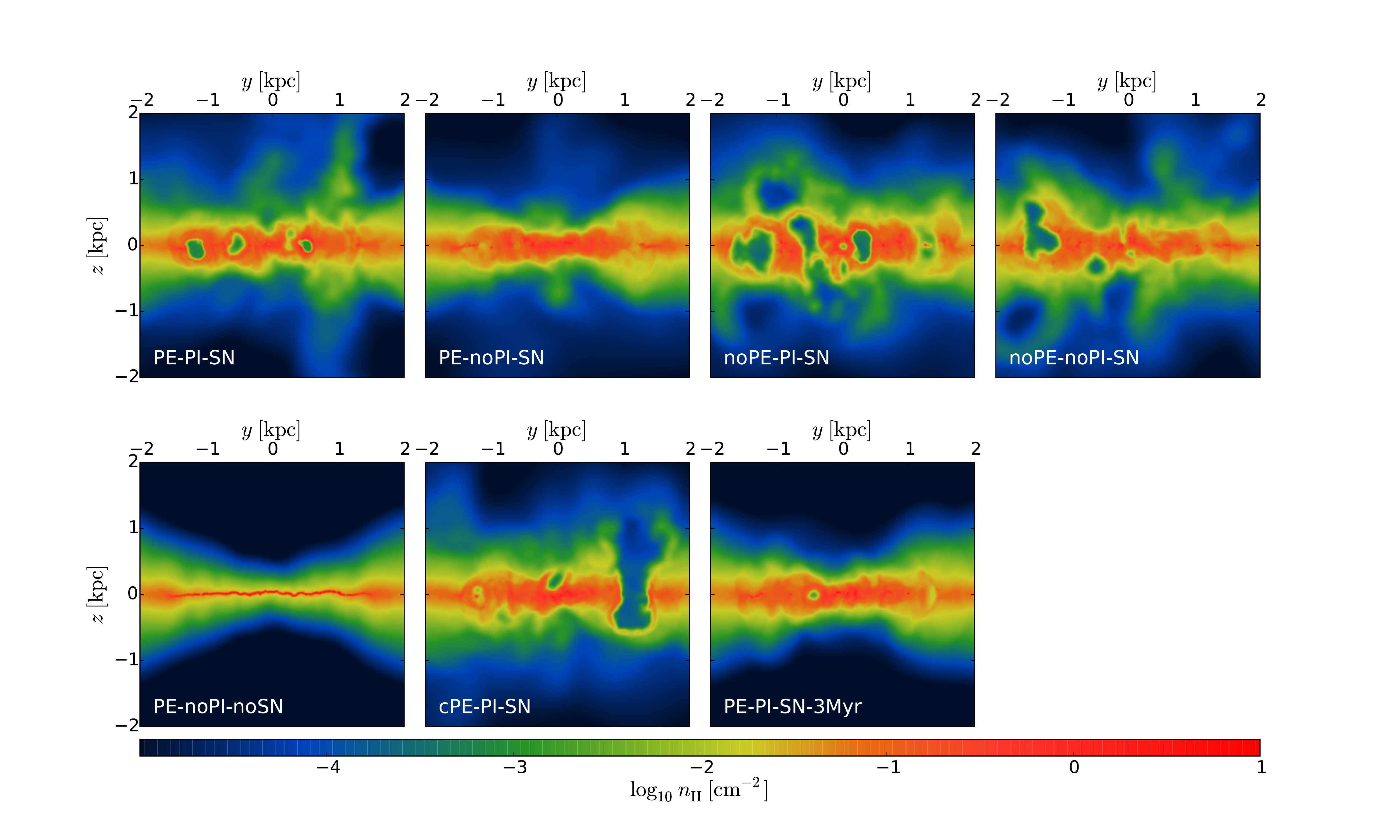}
	\caption{
		Edge-on maps of the hydrogen nuclei number density of all runs for the fiducial disk at $t$ = 500 Myr. 
		The PE-only run (\textit{PE-noPI-noSN}) has a significantly thinner disk compared to the other runs,
		where the disk is puffed up by the turbulent motions triggered by SNe.		
		A quantitative comparison of the (time-averaged) scale-height of the gas disk will be shown in Fig. \ref{fig:sh8runs}.	 
		The feedback-driven bubbles can reach high latitude and push the gas out of the disk.}
	\label{fig:edgeOnMaps500Myr}
\end{figure*}

\subsection{Gas morphology}

Fig. \ref{fig:faceOnMaps500Myr} shows the face-on maps of the hydrogen nuclei column density of all runs for the fiducial disk at $t$ = 500 Myr.
All runs except for the PE-only one (\textit{PE-noPI-noSN}) have similar morphologies: 
a structured ISM with ``bubbles'' driven by vigorous stellar feedback.
\textit{PE-noPI-noSN} appears strikingly different from all other runs,
where the ISM  is clumpier and less dynamic, and it shows little sign of disturbance from stellar feedback.
This implies that PE heating is not energetic enough to trigger blast waves and inject turbulence.
As shown in Fig. \ref{fig:PD_eq_5},
even in the extreme conditions with very strong ISRF,
the dense gas will not be heated above a few hundred of Kelvin as $T$ scales sub-linearly with $G_0$.
This is very different from photoionization which can easily heat up the gas to about $10^4$ K and creates a large pressure imbalance which will subsequently drive an expansion wave.
PE heating is locally a mild feedback process and thus the ISM appears more static in \textit{PE-noPI-noSN}.

Among the other six runs with SNe,
the feedback-driven bubbles are notably smaller in both \textit{PE-noPI-SN} and \textit{PE-PI-SN-3Myr}.
As will be discussed in Section \ref{sec:outflows},
this is because in these two runs very few SNe go off at low densities. 
The SNe therefore become inefficient due to radiative energy loss and therefore very few SN-driven bubbles can form,
even if the star formation rate is very similar to that in all of the other runs (see Fig. \ref{fig:sfr_vs_time}).

Fig. \ref{fig:edgeOnMaps500Myr} shows the edge-on maps of the hydrogen nuclei number density of all runs for the fiducial disk at $t$ = 500 Myr.
The PE-only run (\textit{PE-noPI-noSN}) has a significantly thinner disk compared to other SN runs,
where the disk is puffed up by the turbulent motions triggered by SNe.		
A quantitative comparison of the (time-averaged) scale-height of the gas disk will be shown in Fig. \ref{fig:sh8runs}.
SNe create bubbles that can reach high latitude and push the gas out of the disk.
Again, the bubbles in both \textit{PE-noPI-SN} and \textit{PE-PI-SN-3Myr} are smaller compared to the other runs,
as a result of having most SNe occurring at high densities.

\subsection{Global properties}
\subsubsection{Total star formation rate}

\begin{figure}
	\centering
	\includegraphics[trim = 0mm 0mm 0mm 20mm, clip, width=1\linewidth]{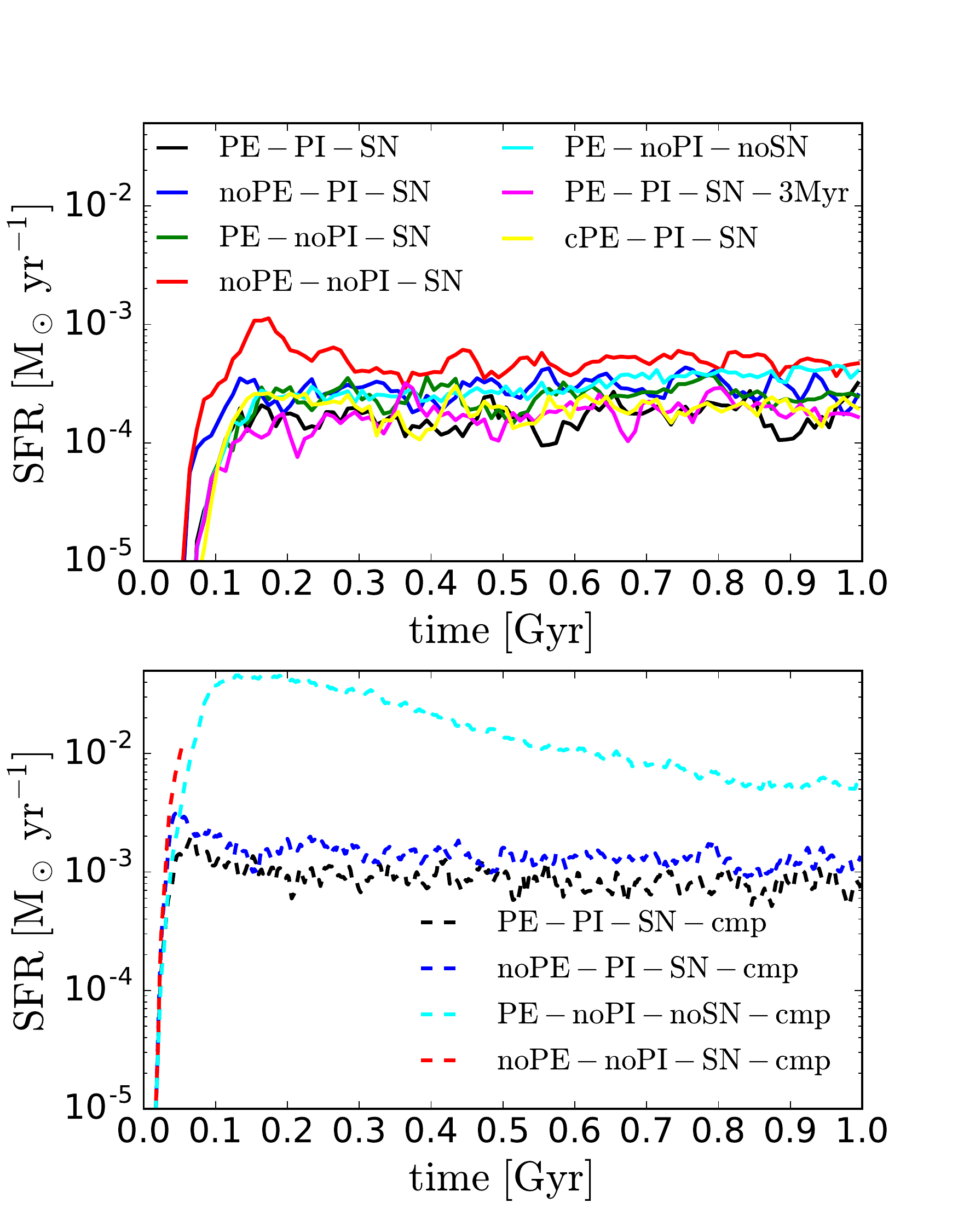}
	\caption{
		Time evolution of the total star formation rate in the fiducial (upper panel) and compact (lower panel) disk.
		All combinations of feedback are able to suppress star formation in the fiducial disk,
		including the PE-only and SN-only runs.
		In the compact disk,
		star formation can only be suppressed when both photoionization and SNe are included.
	}
	\label{fig:sfr_vs_time}
\end{figure}

\begin{figure}
	\centering
	\includegraphics[trim = 50mm 0mm 30mm 0mm, clip, width=0.99\linewidth]{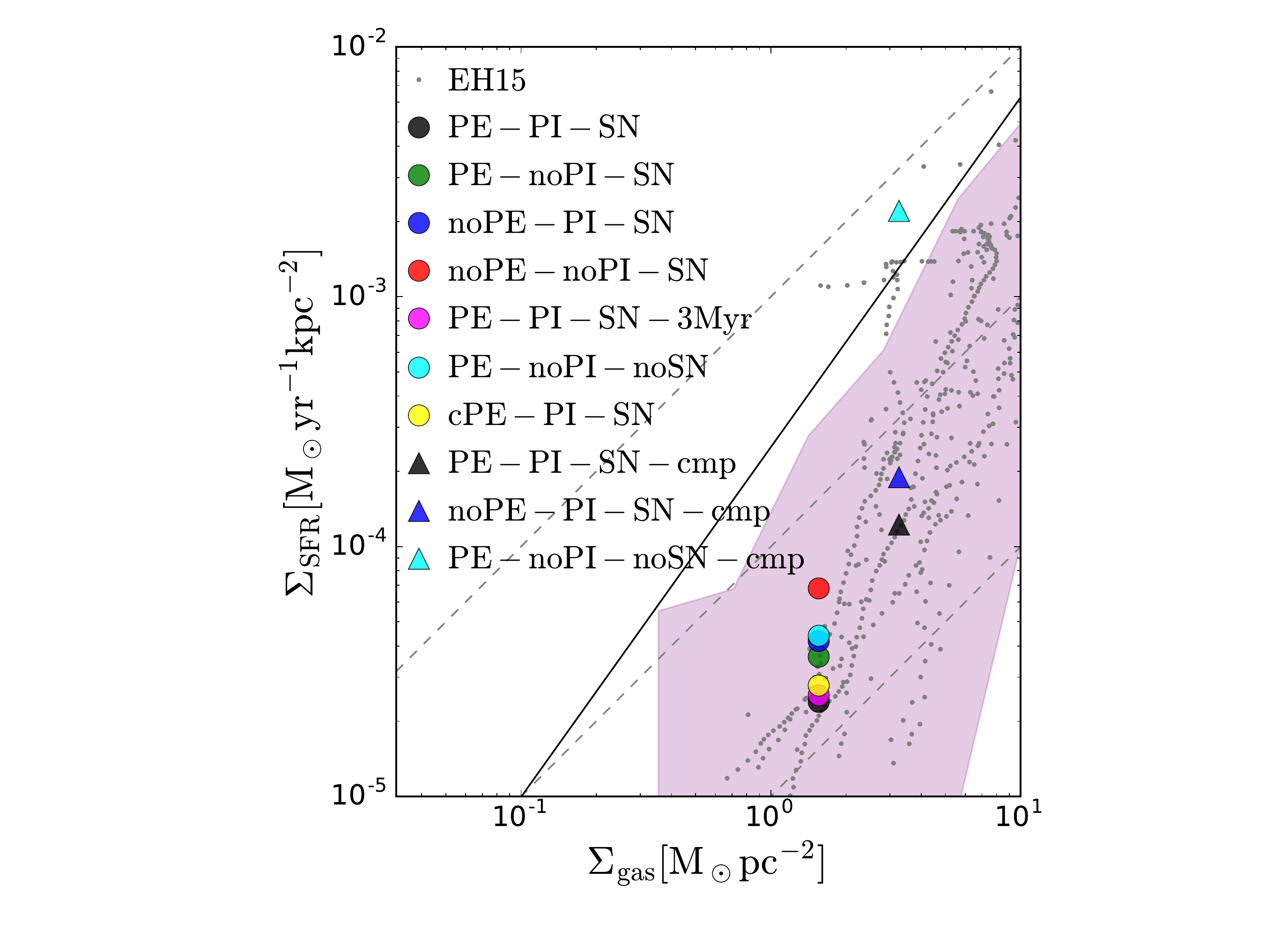}
	\caption{
		Comparison between the simulated galaxies and the observed dwarf galaxies on the Kennicutt-Schmidt plane \citep{1998ApJ...498..541K}: the star formation surface density ($\Sigma_{\rm SFR}$) vs. the gas surface density($\Sigma_{\rm gas}$).
		The gray dots are observed data from \citet{2015ApJ...805..145E} while 
		the purple filled area encloses the fifth and ninety-fifth percentile of the observed data from \citet{2015MNRAS.449.3700R}.
		The solid line shows the conventional Kennicutt-Schmidt relation \citep{1998ApJ...498..541K} while the dashed lines represent the gas depletion times of 1 Gyr, 10Gyr and 100 Gyr from top to bottom.
		All of our models except for \textit{PE-noPI-noSN-cmp} show a reasonable amount of star formation for galaxies in this regime.}
	\label{fig:ksplot}
\end{figure}

\begin{figure}
	\centering
	\includegraphics[trim = 0mm 10mm 0mm 0mm, clip, width=0.99\linewidth]{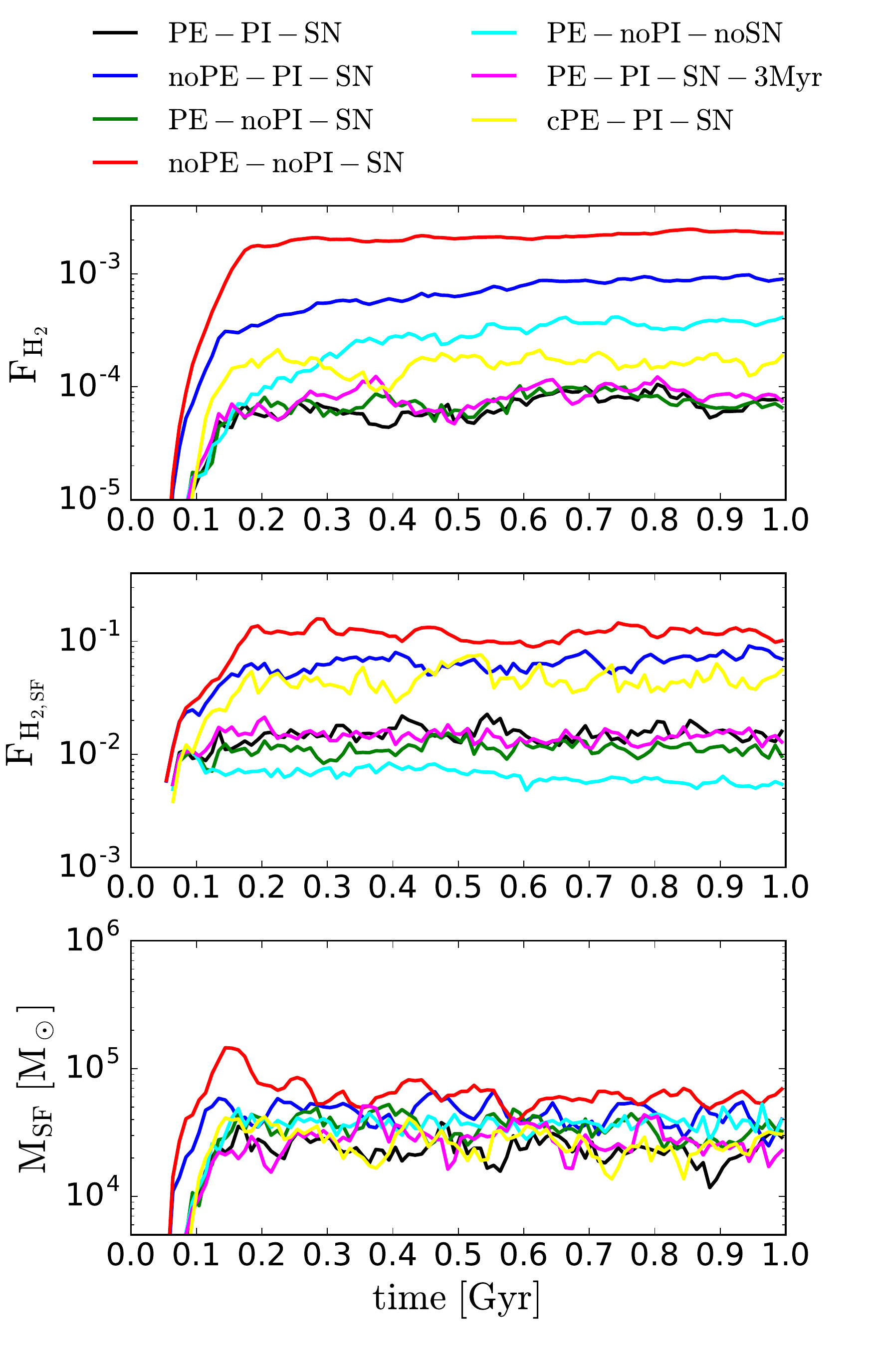}
	\caption{
		Top and middle panels: time evolution of the H$_2$ mass fraction of the total gas (top) and the star-forming gas (middle) in the ISM region ($R <$ 1.5 kpc and $|z| < $ 1 kpc).
		Bottom panel: time evolution of the total mass of the star-forming gas.
		The H$_2$ fraction increases drastically when the ISRF is turned off, suggesting that photodissociation is the main destruction mechanism rather than collisional dissociation triggered by SNe.
		Without SNe, however, most of the gas collapses into dense clumps which also leads to higher H$_2$ fraction.
		Star formation is mainly controlled by the amount of gas that can condense into the cold and dense phase,
		which is set by the competition between gravity and stellar feedback.
	}
	\label{fig:h2_vs_time}
\end{figure}

Fig. \ref{fig:sfr_vs_time} shows the total star formation rate (SFR) of the entire galaxy as a function of time.
For the fiducial disk (solid lines, upper panel),
the total SFR is not very sensitive to any of the feedback combinations that we explore
(their SFRs differ within a factor of two).
A slight trend can still be seen: the more feedback processes we include, the lower is the SFR.
\textit{noPE-noPI-SN} has the highest SFR overall, which reaches a peak of $\sim 10^{-3} {\rm M_\odot\ yr^{-1}}$ in the initial phase ($t <$ 0.2 Gyr),
but then drops by a factor of two afterwards and becomes only slightly higher than the other runs,
This suggests that SNe alone are able to self-regulate the ISM.
Switching on either photoionization (\textit{noPE-PI-SN}) or PE heating (\textit{PE-noPI-SN}) further decreases the SFR,
and the SFR becomes the lowest in the ``full-feedback'' run (\textit{PE-PI-SN}).
Stellar lifetime and the distribution of $G_0$ appear to have negligible effect on the SFR (comparing \textit{PE-PI-SN-3Myr}, \textit{cPE-PI-SN} and \textit{PE-PI-SN}).
The SFR in \textit{PE-noPI-noSN} is similar to \textit{PE-PI-SN} at first but it shows a gradual increase over time and at the end of the simulation ($t$ = 1 Gyr) it is almost as high as the SFR in \textit{noPE-noPI-SN}.

The situation becomes very different in the compact disk (lower panel, dashed lines).
\textit{PE-noPI-noSN-cmp} shows more than an order of magnitude higher SFR than the other two runs,
suggesting that PE heating is no longer able to suppress cooling and gravitational collapse at high gas surface densities.
SNe alone (\textit{noPE-noPI-SN-cmp}) also cannot suppress star formation in the compact disk.
Moreover,
the large amount of unresolved dense gas in this run will later be heated to very high temperatures by SNe,
which leads to timesteps too small for us to continue the simulation after $t > $ 70 Myr,
As this is too short for us to draw any robust conclusion,
we will not discuss \textit{noPE-noPI-SN-cmp} further in this paper.
Including photoionization helps suppress star formation,
as \textit{PE-PI-SN-cmp} and \textit{noPE-PI-SN-cmp} both show much lower SFRs.
This highlights the importance of feedback from massive stars prior to their SN events on regulating star formation in the higher gas surface density regime (see also \citet{2016arXiv160605346G} who investigated the importance of stellar winds in a Milky Way setup).
The fact that \textit{PE-PI-SN-cmp} and \textit{noPE-PI-SN-cmp} both have very similar SFR is consistent with our previous study in Paper I where we found that the SFR in dwarf galaxies is insensitive to the adopted value of $G_0$ (constant field).
The two full-feedback runs in different initial conditions (\textit{PE-PI-SN} and \textit{PE-PI-SN-cmp}) have their SFRs differ by almost an order of magnitude due to the difference in their gas surface density:
SFR scales super-linearly with the gas surface density in this regime \citep{2010AJ....140.1194B, 2011AJ....142...37S}.

Fig. \ref{fig:ksplot} shows the comparison between the simulated galaxies and the observed dwarf galaxies on the Kennicutt-Schmidt plane \citep{1998ApJ...498..541K}: the star formation surface density ($\Sigma_{\rm SFR}$) vs. the gas surface density($\Sigma_{\rm gas}$).
The gray dots are observed data from \citet{2015ApJ...805..145E} while 
the purple filled area encloses the fifth and ninety-fifth percentile of the observed data from \citet{2015MNRAS.449.3700R}.
The total $\Sigma_{\rm SFR}$ and $\Sigma_{\rm gas}$ of the simulated galaxies are calculated
by time-averaging over a time period from $t$=200 Myr to $t$=1000 Myr with a time interval of 10 Myr and assuming the star-forming area is the circular area within $R < $ 1.5 kpc.
The solid line shows the conventional Kennicutt-Schmidt relation \citep{1998ApJ...498..541K} while the dashed lines represent the gas depletion times of 1 Gyr, 10 Gyr and 100 Gyr from top to bottom.
All of our models except for \textit{PE-noPI-noSN-cmp} show a reasonable amount of star formation for galaxies in this regime\footnote{
The data from \citet{2015ApJ...805..145E} and \citet{2015MNRAS.449.3700R} we compare with are in fact both spatially resolved quantities.
As our purpose here is to verify whether the simulated galaxies have a long gas depletion time as seen in observations,
we only present the disk-integrated quantities of the simulated galaxies rather than the pixelwise quantities as we did in Paper I,
which should be sufficient for our purpose.}.

\subsubsection{H$_2$ fraction}

The top panel of Fig. \ref{fig:h2_vs_time} shows the time evolution of the total H$_2$ mass fraction 
(ratio of the mass in H$_2$ to the total hydrogen mass) in the ISM region.
The three runs with both variable PE heating and SNe all show comparable low H$_2$ fractions,
and the \textit{cPE-PI-SN} run has a slightly higher H$_2$ fraction due to its lack of strong radiation field in the vicinity of young stars which can efficiently destroy H$_2$.
Once PE heating is switched off,
the H$_2$ fraction increases drastically by more than one order of magnitude. However, this is an artificial consequence of our implementation: we switch off PE heating by setting $G_{0} = 0$, but this also disables H$_{2}$ photodissociation, and it is the lack of this process that leads to the large change in the H$_{2}$ fraction.
This demonstrates that the collisional dissociation from SNe is a sub-dominant destruction mechanism for H$_2$ compared to photodissociation.
The \textit{noPE-PI-SN} run has a lower H$_2$ fraction than \textit{noPE-noPI-SN} due to our treatment of photoionization, which destroys all H$_2$ within the affected radius.
On the other hand,
the \textit{PE-noPI-noSN} run also has a high H$_2$ fraction even with the ISRF turned on because there is much more cold and dense gas in this run (see Section \ref{sec:PD}).
The fact that all runs shown in Fig. \ref{fig:h2_vs_time} have very different H$_2$ fractions but almost the same SFR demonstrates that H$_2$ does not trace star formation in dwarf galaxies as it does in spiral galaxies, consistent with our findings in Paper I and with prior theoretical predictions \citep{2012MNRAS.426..377G,2012ApJ...759....9K}. Instead,
star formation is mainly controlled by the amount of gas that can condense into the cold and dense phase,
which is set by the competition between gravity and stellar feedback.
The bottom panel of Fig. \ref{fig:h2_vs_time} shows the time evolution of the total mass of the star-forming gas ($M_{\rm SF}$),
which has a very similar behavior to the time evolution of the SFR (see Fig. \ref{fig:sfr_vs_time}).

The middle panel of Fig. \ref{fig:h2_vs_time} shows the time evolution of the H$_2$ mass fraction in the star-forming gas.
In all runs including those with the ISRF switched off, the star-forming gas is dominated by the atomic hydrogen instead of H$_2$.
The ranking of different runs remains in the same order except for \textit{PE-noPI-noSN},
which is related to the fact that  \textit{PE-noPI-noSN} has a very different morphological structure (see Fig. \ref{fig:faceOnMaps500Myr}):
there are many more dense clumps compared to the other runs which leads to a higher total H$_2$ fraction.
However,
these dense clumps are located very close to the UV-emitting stars (see Fig. \ref{fig:PD_3by2}) and therefore
are illuminated by a much stronger ISRF which leads to the lowest H$_2$ fraction in star-forming gas in all runs.

\subsection{Local properties}\label{sec:PD}


\begin{figure}
	\centering
	\includegraphics[trim = 10mm 10mm 0mm 0mm, clip, width=1.\linewidth]{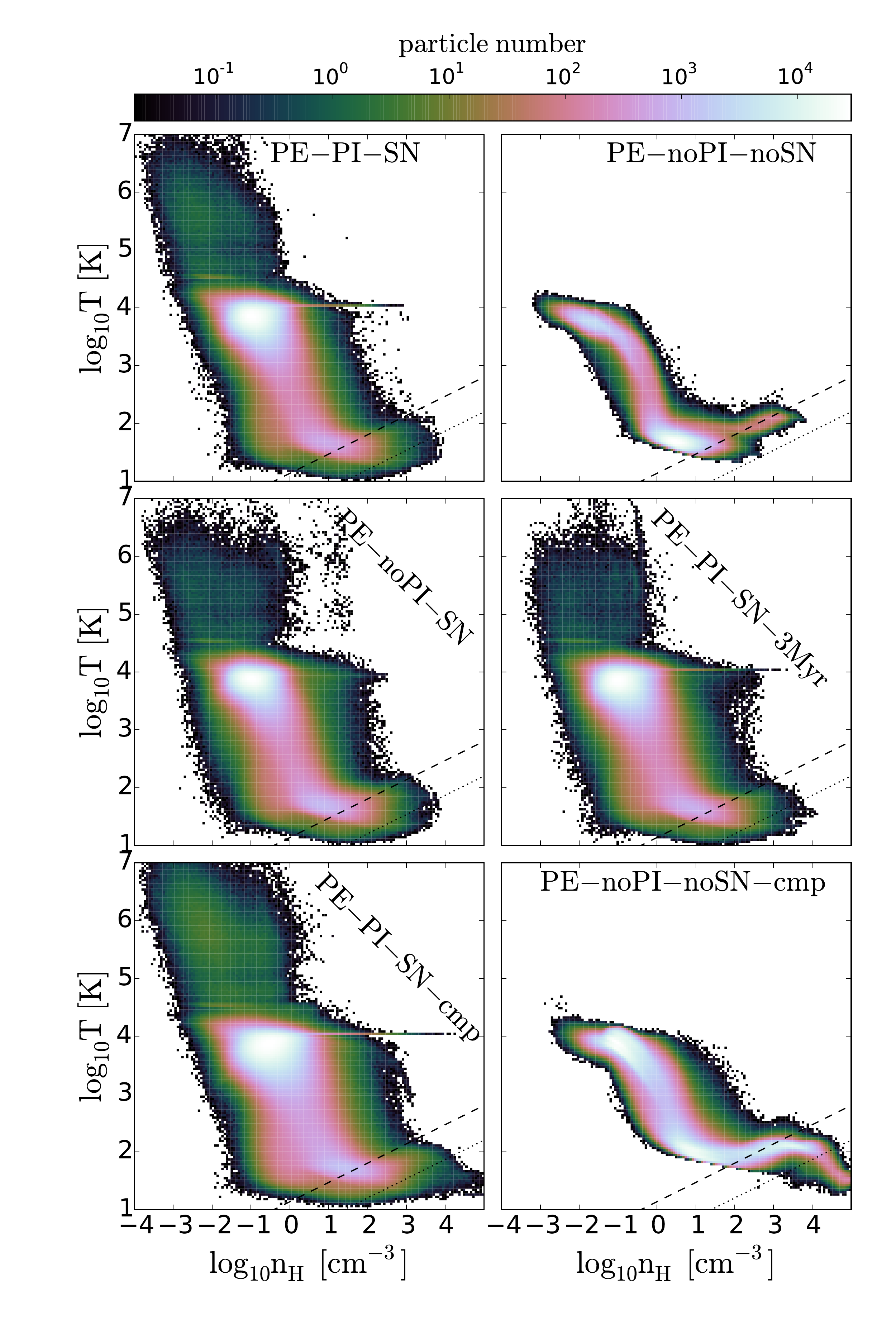}
	\caption{
		Phase diagrams (temperature vs. number density) for the gas in the ISM region in all the runs that include self-consistent PE heating, 
		averaged over a period from $t$ = 400 Myr to $t$ = 900 Myr with a time interval of 10 Myr. 
		The PE-only runs show significantly different distributions from the other runs:
		they have much more cold gas and have no hot gas at all, and the star-forming gas is notably warmer than in the other runs. The dashed lines represent our star formation threshold where $M_{\rm J} = 8 M_{\rm ker}$ while the dotted lines represent the resolution limit where $M_{\rm J} = M_{\rm ker}$.}
	\label{fig:PD_3by2}
\end{figure}

\begin{figure}
	\centering
	\includegraphics[trim = 10mm 10mm 0mm 0mm, clip, width=1.\linewidth]{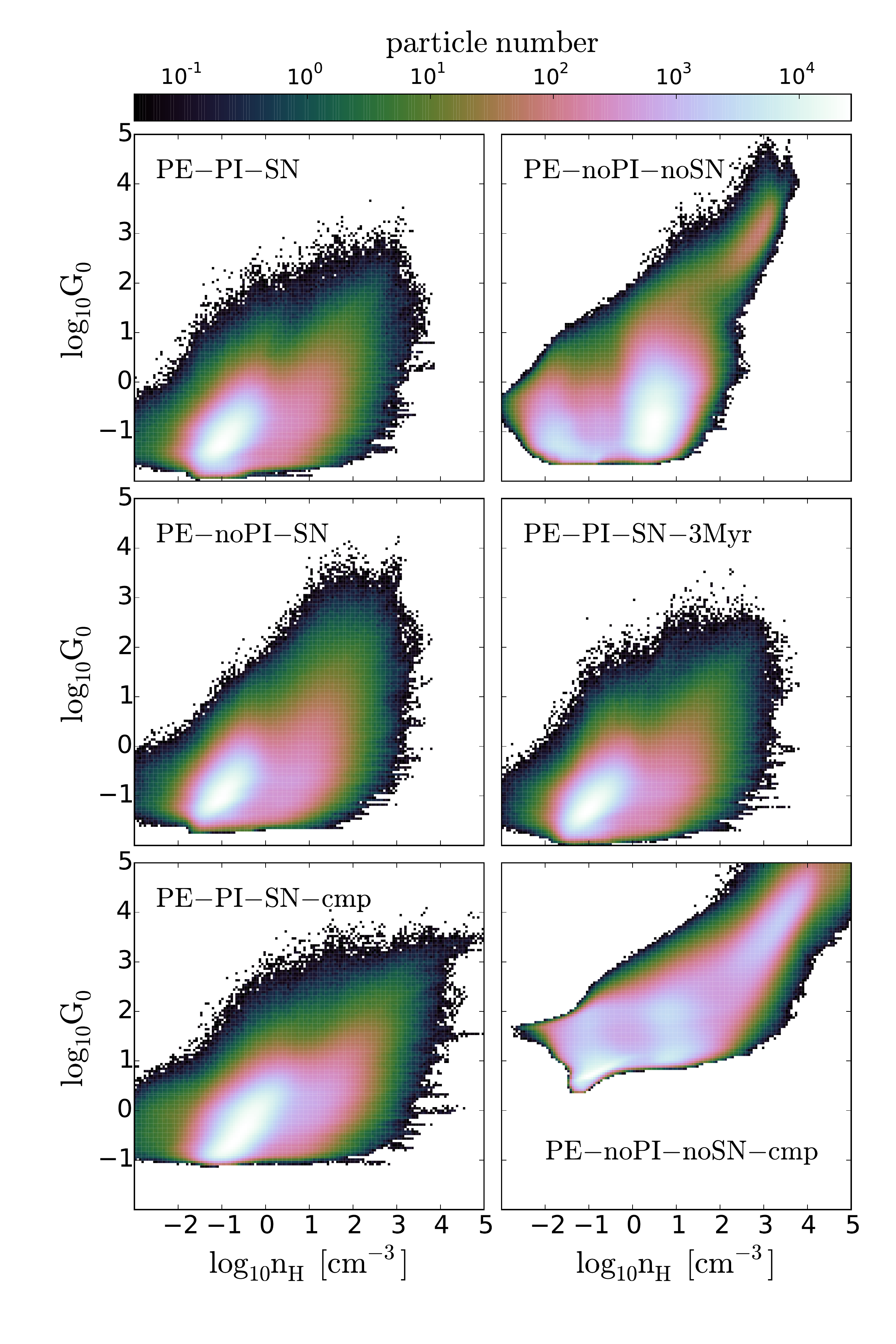}
	\caption{
		The strength of the ISRF ($G_0$) vs. the hydrogen number density ($n_{\rm H}$) for gas in the ISM region,
		averaged over a period from $t$ = 400 Myr to $t$ = 900 Myr with a time interval of 10 Myr. 
		Dense gas in the runs with PE heating alone is subject to much higher $G_0$ as the dense clumps are much closer to the radiation sources. 
		With supernovae the gas is dispersed and so  $G_0$ is much lower.}
	\label{fig:rho_vs_G0}
\end{figure}

Fig. \ref{fig:PD_3by2} shows the gas distribution on the phase diagram (number density vs. temperature) in the ISM region in all six runs that include self-consistent PE heating,
averaged over a period from $t$ = 400 Myr to $t$ = 900 Myr with a time interval of 10 Myr.
The dotted lines show the contours where $M_{\rm J} = M_{\rm ker} = 400 \, {\rm M_\odot}$,
which can be viewed as our resolution limit.
The dashed lines show the star formation threshold, $M_{\rm J} = 8 M_{\rm ker}$.
The two runs with only PE heating (\textit{PE-noPI-noSN} and \textit{PE-noPI-noSN-cmp}) show very different distributions from the other runs:
there is much more cold gas when PE heating is the only feedback mechanism.
In the fiducial disk (\textit{PE-noPI-noSN}),
PE heating alone is able to heat up the dense gas to $T \gtrsim 100$ K,
which appears to be enough to stop further gravitational collapse and therefore maintain a low SFR.
However, for the compact disk (\textit{PE-noPI-noSN-cmp}),
a significant amount of gas collapses to very high densities ($n \gtrsim 10^3 - 10^4$ cm$^{-3}$) even though the gas is also heated up to $T \gtrsim 100$ K by PE heating, 
and thus the SFR is boosted dramatically (see Fig. \ref{fig:sfr_vs_time}).

The other four runs with SNe in Fig. \ref{fig:PD_3by2}  have a more scattered distribution,
mainly due to enhanced turbulent motions driven by SNe.
These runs also have hot gas ($T > 3\times 10^4$ K) which can only be generated by SNe.
The dense gas in these runs has a slightly cooler gas temperature on average (below 100 K).
The narrow lines at $T = 10^4$ K are due to photoionization.
The majority of gas is warm ($T\sim 10^4$ K). Notably, it is much warmer than its thermal equilibrium temperature (shown in Fig. \ref{fig:PD_eq_5}),
as the gas is constantly stirred and shocked by both SNe and photoionization and has insufficient time to cool.
As shown in Paper I, this occurs when the gas cooling time is longer than the turbulent dynamical time,
which is more likely to be the case in low-metallicity systems.
In \textit{PE-PI-SN-cmp}, the gas retained at $10^4$ K extends to slightly higher densities compared to the other three runs because of the higher SFR and therefore higher SN rate.

\begin{figure}
	\centering
	\includegraphics[trim = 20mm 5mm 0mm 0mm, clip, width=1\linewidth]{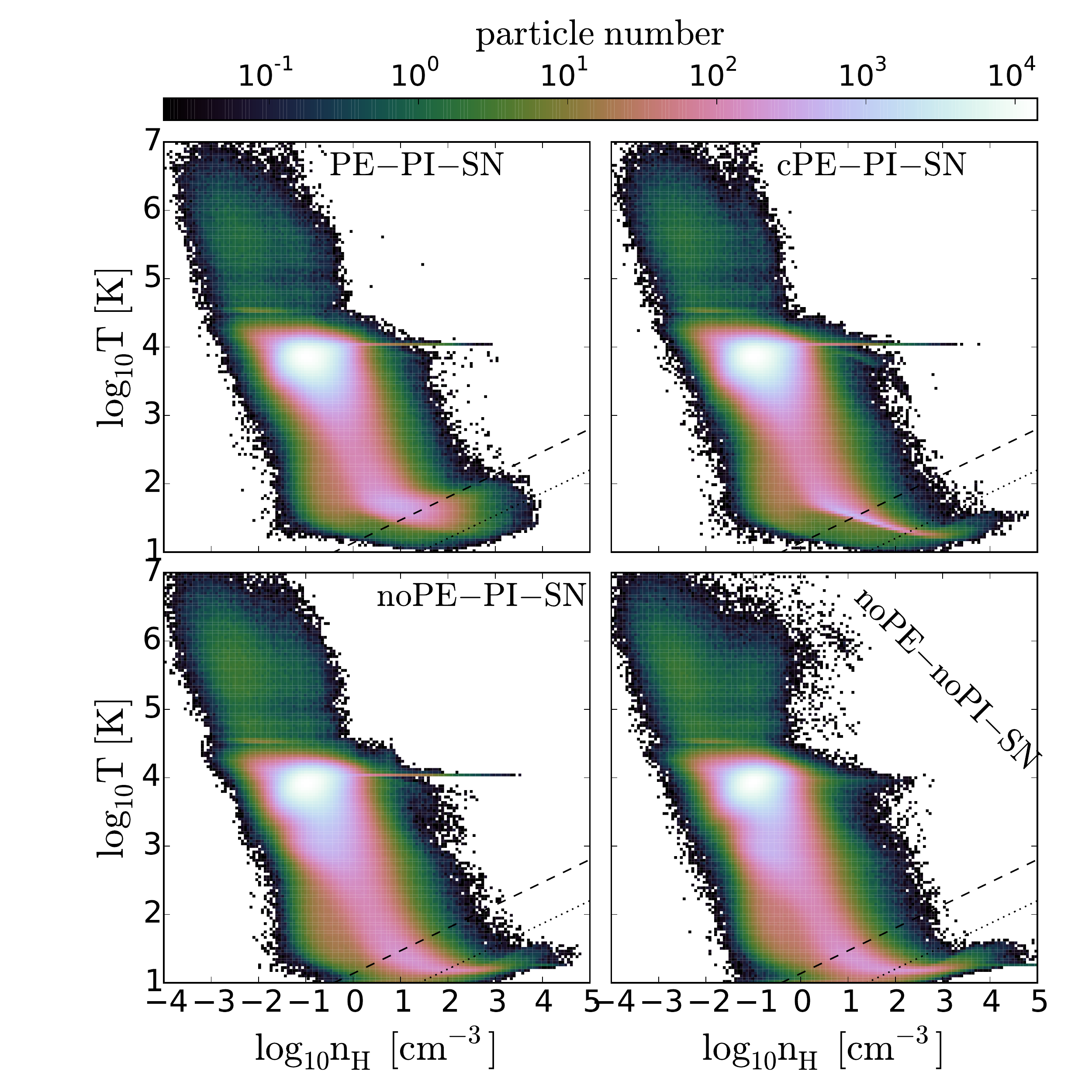}
	\caption{
		Phase diagrams of the ISM in the runs with variable $G_0$ (upper left), 
		constant $G_0$ (upper right) and $G_0$ = 0	(lower panels),
		averaged over a period from $t$ = 400 Myr to $t$ = 900 Myr with a time interval of 10 Myr. 
		Without variable PE heating, the cold gas is slightly colder, but the overall distribution does not change much.
		SNe are able to keep the majority of gas warm ($\sim 10^4$ K) even without PE heating, which is consistent with our findings in Paper I.}
	\label{fig:PD_2by2}
\end{figure}

Fig. \ref{fig:rho_vs_G0} shows the ISM distribution on a $n$-$G_0$ plane (number density vs. ISRF) for the same set of runs as in Fig. \ref{fig:PD_3by2},
averaged over the same time period from $t$ = 400 Myr to $t$ = 900 Myr with a time interval of 10 Myr.
In the two PE-only runs (\textit{PE-noPI-noSN} and \textit{PE-noPI-noSN-cmp}),
the dense gas is illuminated by a very strong ISRF with orders of magnitude higher $G_0$  compared to that in the other four runs. This is responsible for the much higher temperatures found in the star-forming gas in these runs (see  Fig. \ref{fig:PD_3by2}).
The lack of SNe (and/or photoionzation) results in a less turbulent ISM where most of the gas will simply be attracted towards the star-forming clumps due to gravity.
Therefore,
the gas experiences a much higher $G_0$ due to its proximity to the UV-emitting stars.
On the other hand,
once SNe are included,
they trigger blastwaves and turbulence which disperse the nearby clouds.
As such,
the dense gas on average is further away from the sources, and so the typical $G_{0}$ found in the dense gas drops significantly.
This means that PE heating in these runs is actually much weaker than it is in the PE-only runs,
where the omission of other feedback mechanisms (SNe and photoionization) turns out to significantly boost the effect of PE heating.
This effect is mainly caused by a change in the gas distribution and not by the energy budget of the ISRF (as the total SFR is almost unchanged).


Fig. \ref{fig:PD_2by2} shows the phase diagrams of the ISM in the runs with variable $G_0$ (upper left), constant $G_0$ (upper right) and $G_0$ = 0	(lower panels),
averaged over a period from $t$ = 400 Myr to $t$ = 900 Myr with a time interval of 10 Myr.
The gas distribution is similar in all of these runs, with the majority of the gas being warm ($T \sim 10^4$ K) and diffuse ($n \sim 0.1$ cm$^{-3}$).
The main difference is that without PE heating (\textit{noPE-PI-SN} and \textit{noPE-noPI-SN}) the cold gas can cool down to slightly lower temperatures (by $\sim$ 0.5 dex).
This makes the system more gravitationally unstable and results in more star-forming gas,
which manifests as a ``tail'' in the phase diagram that further extends into the dense unresolved regions (to the right of the dotted line).
Switching on the (variable) PE heating (\textit{PE-PI-SN}) makes most of the star-forming gas stay at the resolved region.
This difference in the star-forming gas leads to the slightly higher total SFR in Fig. \ref{fig:sfr_vs_time} for the no PE heating runs.
The constant-$G_0$ run (\textit{cPE-PI-SN}) shows a behavior intermediate between the variable-$G_0$ run and the no-$G_0$ run:
it has a slightly warmer cold gas due to the PE heating from the background radiation field,
but it also has a high-density tail as there is no enhanced PE heating in the densest gas.
In all cases, SNe are able to keep the majority of gas warm ($\sim 10^4$ K) even in the SN-only run (\textit{noPE-noPI-SN}).

\subsection{FIR lines and IR continuum emission}

\begin{table*}
	\caption{The time-averaged (over a time period from $t$=200 Myr to $t$=1000 Myr) total luminosities of the continuum dust emission ($L_{\rm cont} [L_\odot]$) and FIR lines ($L_{\rm [CII]157}, L_{\rm [OI]63}$ and $L_{\rm [OI]145}$) as well as their fluctuations in all runs.}
	\label{table:lines_and_dust}
	\begin{tabular}{| l | l | l | l | l |}    
		\hline\hline
		Name                   & $L_{\rm cont} [L_\odot]$ & $L_{\rm [CII]157} [L_\odot]$ & $L_{\rm [OI]63} [L_\odot]$  & $L_{\rm [OI]145} [L_\odot]$ \\
		\hline
		\textit{PE-PI-SN}            & $(1.599\pm 0.407) \times 10^5$ & $(613.6\pm 149.4)$  & $(771.2\pm 148.3)$ & $(72.47\pm 16.67)$ \\
		\textit{PE-noPI-SN}       & $(4.002\pm 1.481) \times 10^5$ &  $(571.8\pm 76.1) $    & $(638.8\pm 61.9)$  & $(56.75\pm 5.39)$ \\
		\textit{noPE-PI-SN}       & $(23.11\pm 4.04)  $                      & $(633.1\pm 216.1) $  & $(861.5\pm 257.1)$ & $(83.13\pm 28.46)$ \\
		\textit{noPE-noPI-SN}   & $(21.65\pm 15.68)$                      & $(366.8\pm 47.3)$   & $(557.9\pm 77.0)$   & $(50.61\pm 6.81)$ \\
		\textit{PE-PI-SN-3Myr}  & $(1.341\pm 0.409) \times 10^5$& $(575.6\pm 152.2) $  & $(762.2\pm 169.6)$& $(71.35\pm 18.69)$ \\
		\textit{PE-noPI-noSN}   & $(1.951\pm 0.557) \times 10^6$ & $(1092\pm 136.7) $ & $(302.9\pm 47.1)$  & $(16.04\pm 5.53)$ \\
		\textit{cPE-PI-SN}          & $(1.144\pm 0.030) \times 10^5$ & $(620.5\pm 222.0) $ & $(828.2\pm 240.4)$ & $(79.38\pm 26.99)$ \\
		\hline
		\textit{PE-PI-SN-cmp}           & $(2.025\pm 0.364) \times 10^6$ & $(5.600\pm 1.211)\times 10^3 $ & $(6.401\pm 1.662)\times 10^3$ & $(623.0\pm 174.7)$ \\
		\textit{noPE-PI-SN-cmp}       & $(120.4\pm 27.8) $                       & $(4.763\pm 1.015)\times 10^3 $ & $(5.391\pm 1.173)\times 10^3$ & $(536.3\pm 124.5)$ \\				
		\textit{PE-noPI-noSN-cmp}  & $(1.252\pm 1.196) \times 10^8$   & $(3.416\pm 2.636)\times 10^4 $  & $(2.182\pm 1.999)\times 10^4$  & $(863.5\pm 821.4)$ \\				
		\hline\hline
	\end{tabular}
\end{table*}

\begin{figure}
	\centering
	\includegraphics[trim = 5mm 10mm 0mm 0mm, clip, width=0.99\linewidth]{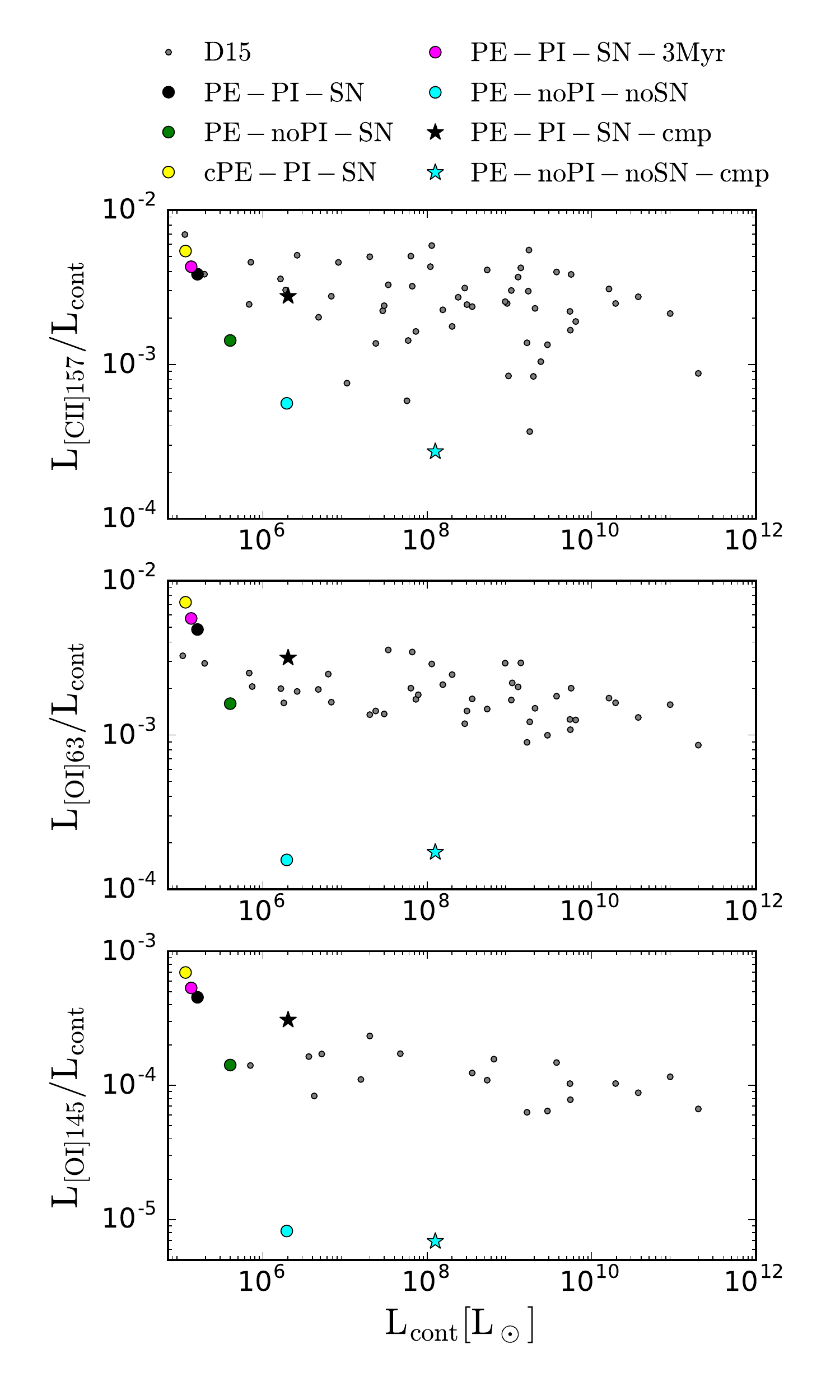}
	\caption{The total luminosity of the continuum infrared emission $L_{\rm cont}$ vs. the ratio of FIR lines to the continuum for all runs with PE heating,
		time-averaged over a time period from $t$=200 Myr to $t$=1000 Myr.
		The observed data from the Dwarf Galaxy Survey from \citet{2015A&A...578A..53C} are over-plotted as gray dots.
		The PE-only runs have their FIR-to-continuum ratios much lower than the observed dwarf galaxies,
		especially for the oxygen lines.
		Including SNe brings the simulated galaxies into locations in the plot that agree well with observations by pushing gas further away from the massive stars.
	}
	\label{fig:compare_DGS}
\end{figure}

\begin{figure*}
	\centering
	\includegraphics[trim =20mm 20mm 0mm 0mm, clip, width=0.99\linewidth]{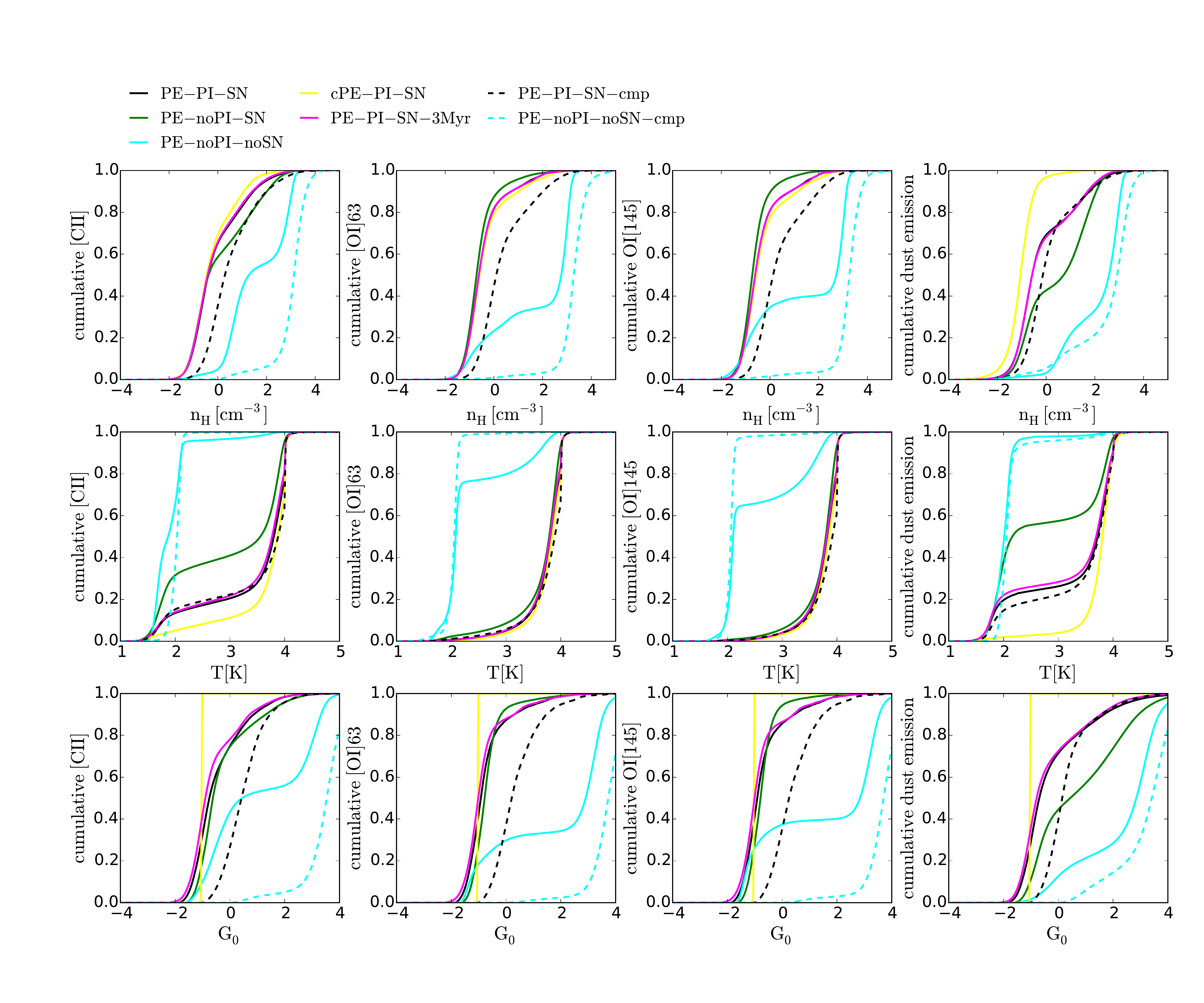}
	\caption{
		Normalized cumulative distribution of the IR continuum emission from dust and the FIR metal lines as a function of hydrogen number density ($n_{\rm H}$, top row), gas temperature ($T$, middle row) and local radiation field ($G_0$, bottom row).
		Runs with SNe have their FIR metal lines primarily coming from the diffuse ($n_{\rm H} <$ 1 cm$^{-3}$) and warm ($T > 3\times10^3$ K) phase of the ISM.
		On the other hand,
		due to the very different gas morphology in the PE-only runs (\textit{PE-noPI-noSN} and \textit{PE-noPI-noSN-cmp}),
		most of the FIR lines and the continuum are coming from the cold and dense gas exposed to a strong ISRF in these runs.
	}
	\label{fig:obs_cumu}
\end{figure*}

We estimate the total luminosity from the far infrared (FIR) fine structure emission lines [CII]157$\mu$m, [OI]63$\mu$m and [OI]145$\mu$m for the simulated galaxies via post-processing, denoted as $L_{\rm [CII]157}, L_{\rm [OI]63}, L_{\rm [OI]145}$, respectively.
The emissivity of the FIR lines for each gas particle is obtained using our non-equilibrium cooling module albeit assuming optically thin conditions,
which is a fair approximation for low-metallicity gas\footnote{The optically thin approximation breaks down at high column densities. For example, for [CII]157$\mu$m this occurs at a total hydrogen column density $N_{\rm H} \sim 10^{22} \, {\rm cm}^{-2}$ given our $0.1 \, Z_{\odot}$ metallicity and an assumed [CII] line-width of $2.4 \, {\rm km \, s^{-1}}$, and at a proportionately higher column density if the line-width is larger than this value. A similar column density is required to produce $\tau \sim 1$ in the  [OI]63$\mu$m line \citep{2016MNRAS.462.3011G}. These values are ten times higher than the column densities found in our setups, even in our compact disk simulations.  In the very dense gas where this approximation does break down we will overestimate the luminosity, but this will only affect a very small fraction of the gas in our simulations.}.
We then obtain the total luminosity by summing over the contributions of all gas particles within the circular area $R <$ 1.5 kpc.

In addition, we estimate the continuum infrared emission from dust.
The local dust temperature in our simulations is mainly dictated by dust heating due to the local ISRF and the radiative cooling of dust,
the latter of which is obtained using the fitting formula given by \citet{2012MNRAS.421..116G}:
\begin{equation}
\Lambda_{\rm dust} (T_{\rm d}) = 4.68 \times 10^{-31} D' T_{\rm d}^6 n_{\rm H}\ {\rm erg\ s^{-1}\ cm^{-3}}, \label{eq:dusttemp}
\end{equation}
where $D'$ is the dust-to-gas ratio relative to the solar value,
$T_{\rm d}$ is the dust temperature and $n_{\rm H}$ is the hydrogen nuclei number density.
Assuming optically thin conditions,
the continuum infrared emission ($L_{\rm cont}$) is obtained by summing over the contributions of all gas particles within the circular area $R <$ 1.5 kpc.\footnote{Note that as Equation~\ref{eq:dusttemp} was derived assuming \citet{1994A&A...291..943O} dust opacities, we are implicitly assuming here that in our low metallicity dwarf galaxy, the dependence of opacity on wavelength has the same form, and it is only the normalization that is different owing to the lower dust-to-gas ratio.}

Fig. \ref{fig:compare_DGS} shows the relation between $L_{\rm cont}$ and the ratio of FIR lines to the dust continuum for all runs with PE heating,
averaged over a time period from $t$=200 Myr to $t$=1000 Myr with a time interval of 10 Myr.
The observed data from the Dwarf Galaxy Survey \citep{2015A&A...578A..53C} are over-plotted as gray dots.
Since the \textit{noPE} runs have $G_0 = 0$ everywhere and thus have no dust heating due to the ISRF,
their $L_{\rm cont}$ values are around four orders of magnitude lower than in runs with PE heating and therefore are not visible in Fig. \ref{fig:compare_DGS}.
We report the averaged $L_{\rm cont}$ and the FIR line luminosities as well as their fluctuations in all runs in Table \ref{table:lines_and_dust}.

For the fiducial disk,
\textit{PE-PI-SN} and \textit{PE-PI-SN-3Myr} show similar continuum and FIR line luminosities. This is unsurprising, as their gas distributions in the $n_{\rm H} - G_0$ plane and in the phase diagram are similar. $L_{\rm cont}$ in \textit{PE-noPI-SN} is about 2.5 higher than in \textit{PE-PI-SN} as in the former case there is slightly more dense gas close to the massive stars that experiences a stronger ISRF.
On the other hand, the FIR line luminosities in \textit{PE-noPI-SN} do not differ much from those in \textit{PE-PI-SN},
which implies that FIR lines mainly originate from the diffuse gas (cf. Fig. \ref{fig:obs_cumu}).
In the constant-$G_0$ run (\textit{cPE-PI-SN}),
$L_{\rm cont}$ is about 70\% of that in \textit{PE-PI-SN},
as the former lacks the contributions from the high-$G_0$ regions close to massive stars.
It also has comparable FIR line luminosities to those in \textit{PE-PI-SN},
for the same reason that the FIR lines mainly come from the diffuse gas which does not differ much in these runs.

Most strikingly,
$L_{\rm cont}$ in \textit{PE-noPI-noSN} is an order of magnitude higher than in \textit{PE-PI-SN},
as there is a significant fraction of high-$G_0$ gas in the former case (see Fig. \ref{fig:rho_vs_G0}).
It also has a higher $L_{\rm [CII]157}$ and lower $L_{\rm [OI]63}$ and $L_{\rm [OI]145}$ than in \textit{PE-PI-SN} as a result of being dominated by cold gas rather than by warm gas.
In the compact disk,
the difference in $L_{\rm cont}$ between \textit{PE-PI-SN-cmp} and \textit{PE-noPI-noSN-cmp} is even more pronounced:
it is two orders of magnitude higher in the latter case which has a huge number of dense clumps with high $G_0$.
The FIR-to-continuum ratios in the PE-only runs are much lower than the values observed in real dwarf galaxies, especially for the oxygen lines.
Including SNe brings the simulated galaxies into the locations that agree well with observations by pushing gas further away from the massive stars.

\begin{figure}
	\centering
	\includegraphics[trim = 0mm 0mm 0mm 10mm, clip, width=1.\linewidth]{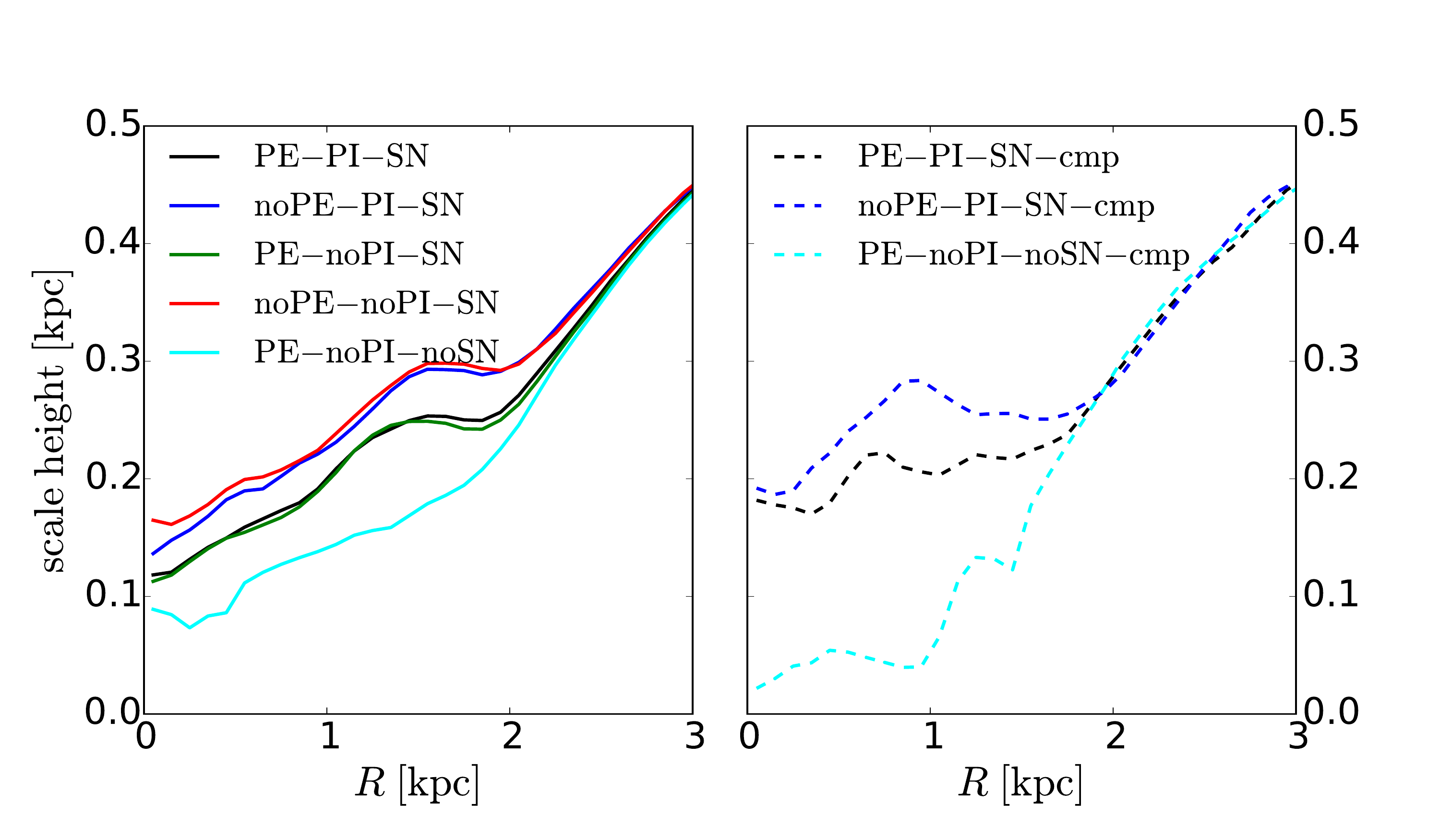}
	\caption{
		The scale-height of the gaseous disk as a function of $R$ in eight different runs,
		averaged over a period from $t$ = 400 Myr to $t$ = 900 Myr with a time interval of 10 Myr.
		The left panel shows the runs for the fiducial disk with a scale-length of the gas disk $l_{\rm gas} = 1.46$ kpc and the right panel for the compact disk with $l_{\rm gas} = 0.73$ kpc.
		The scale-height increases with the SN rate, which scales with the SFR (if SNe are included) shown in Fig. \ref{fig:sfr_vs_time}. 
		The lowest scale-heights are found in the runs without SNe.}
	\label{fig:sh8runs}
\end{figure}

\subsubsection{Cumulative distribution of IR emissions}

Fig. \ref{fig:obs_cumu} shows the normalized cumulative distribution of the IR continuum emission from dust and the FIR metal lines as a function of hydrogen number density ($n_{\rm H}$, top row), gas temperature ($T$, middle row) and local radiation field ($G_0$, bottom row),
averaged over a time period from $t$=200 Myr to $t$=1000 Myr with a time interval of 10 Myr.
Only runs with PE heating are shown.
Runs with SNe, including the full-feedback models (\textit{PE-PI-SN} and \textit{PE-PI-SN-cmp}),
have their FIR metal lines primarily coming from the diffuse ($n_{\rm H} <$ 1 cm$^{-3}$) and warm ($T > 3\times10^3$ K) phase of the ISM.
The ISM in these runs is dominated by the warm and diffuse phase both in volume and in mass,
as will be shown in Fig. \ref{fig:hot_MVfrac}.
The [CII]157$\mu$m line has a slightly higher contribution from the cold ($T < $ 100 K) gas compared to the oxygen lines due to its lower excitation level.
\textit{PE-noPI-SN} has around 20\% higher contribution from the cold phase than the fiducial run \textit{PE-PI-SN} because it has more cold and dense gas which is effectively destroyed by photoionization in the latter case.
As our simple treatment of photoionization is likely to have an excessive effect on destroying dense gas,
we expect that a more realistic ISM should be intermediate between \textit{PE-PI-SN} and \textit{PE-noPI-SN}.
The FIR lines do not trace the regions where the ISRF is strong,
as the majority of the emission is coming from gas with a $G_0$ not much higher than the background value ($G_0 \sim$ 0.1 for the fiducial disk).
The dust continuum emission originates from a slightly denser and cooler gas which experiences a stronger ISRF.
On the other hand,
due to the very different gas morphology in the PE-only runs (\textit{PE-noPI-noSN} and \textit{PE-noPI-noSN-cmp}),
most of the FIR lines and the continuum are coming from the cold and dense gas with a strong ISRF in these runs.
We caution that our simulations do not resolve the internal structure of the dense clouds
and therefore may underestimate the emissions from those regions if there is,
as inferred from observations such as \citet{2015A&A...578A..53C} and \citet{2016AJ....151...14C}, 
significant clumping of gas at the scales below our resolutions. 
However,
we expect that the systematic trends among different runs to be qualitatively unchanged.

\subsection{Scale-height of the gaseous disk}

Fig. \ref{fig:sh8runs} shows the scale-height of the gas disk as a function of $R$ in eight different runs,
averaged over a period from $t$ = 400 Myr to $t$ = 900 Myr with a time interval of 10 Myr. 
The left panel shows the runs for the fiducial disk with a scale-length of the gas disk $l_{\rm gas} = 1.46$ kpc and the right panel for the compact disk with $l_{\rm gas} = 0.73$ kpc.
The scale-height is defined as the latitude $z_s$ where 75 \% of the gas is enclosed between $z = \pm z_s$ at the given radial bin.
A significant flaring can be seen in all cases.
In the left panel,
the \textit{noPE-noPI-SN} and \textit{noPE-PI-SN} runs have the largest scale-height at all radii.
This implies a correlation between gas scale-height and the SN rates (due to their slightly higher SFR)
as the SNe enhance both the thermal (by shock heating) and kinetic (by turbulence) pressure support.
Photoionization seems to plays a minor role in affecting the gas scale-height.
The \textit{PE-noPI-noSN} run has the lowest scale-height because its SN rate is zero even though it has a SFR comparable to the other runs.
PE heating itself does not trigger significant turbulence.
In the right panel, a similar trend can be seen that the scale-height is positively correlated with the SN rate.
In addition,
turning off supernovae in the compact disk has a much stronger effect than in the fiducial disk,
as the vertical gravity is stronger in the former case.
This is connected to the fact that in the absence of SN feedback, star formation can only be  suppressed in the fiducial disk and not in the compact disk (see Fig. \ref{fig:sfr_vs_time}).

\subsection{Galactic Outflows}\label{sec:outflows}

\begin{figure}
	\centering
	\includegraphics[trim = 0mm 10mm 0mm 20mm, clip, width=1\linewidth]{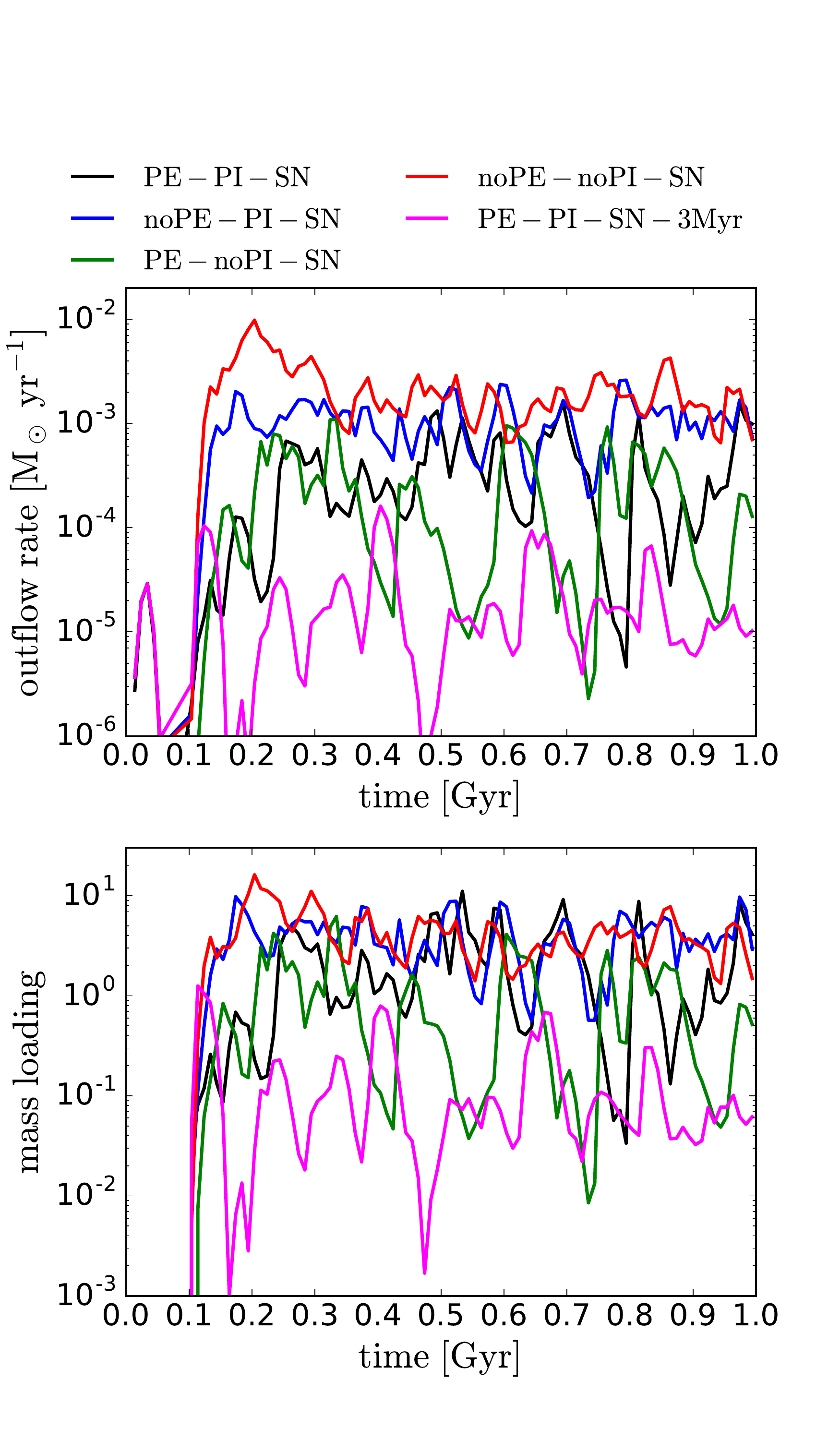}
	\caption{
		Time evolution of the outflow rate (upper panel) and the mass loading factor (the ratio of the outflow rate to the SFR, lower panel), measured at $|z|$ = 2 kpc. 
		The outflow rate is much more sensitive to the adopted feedback compared to the SFR (see Fig. \ref{fig:sfr_vs_time}).}
	\label{fig:outflow_mass_loading_2kpc_time}
\end{figure}

In the upper panel of Fig. \ref{fig:outflow_mass_loading_2kpc_time} we show the galactic outflow rate for the fiducial disk as a function of time.
The outflow rate is defined in the same way as in Paper I,
which measures the mass flux through the two planes at $z = \pm$ 2 kpc along the outflowing directions.
Compared to the SFR, the outflow rate fluctuates much more over time,
and it is quite sensitive to the adopted feedback model.
Run \textit{PE-noPI-noSN} is not shown because its outflow rate is essentially zero.
In the lower panel of Fig. \ref{fig:outflow_mass_loading_2kpc_time} we show the mass loading factor, defined as the ratio of the SFR to the outflow rate,  as a function of time in the same set of runs.

\begin{figure}
	\centering
	\includegraphics[trim = 0mm 0mm 0mm 0mm, clip, width=1\linewidth]{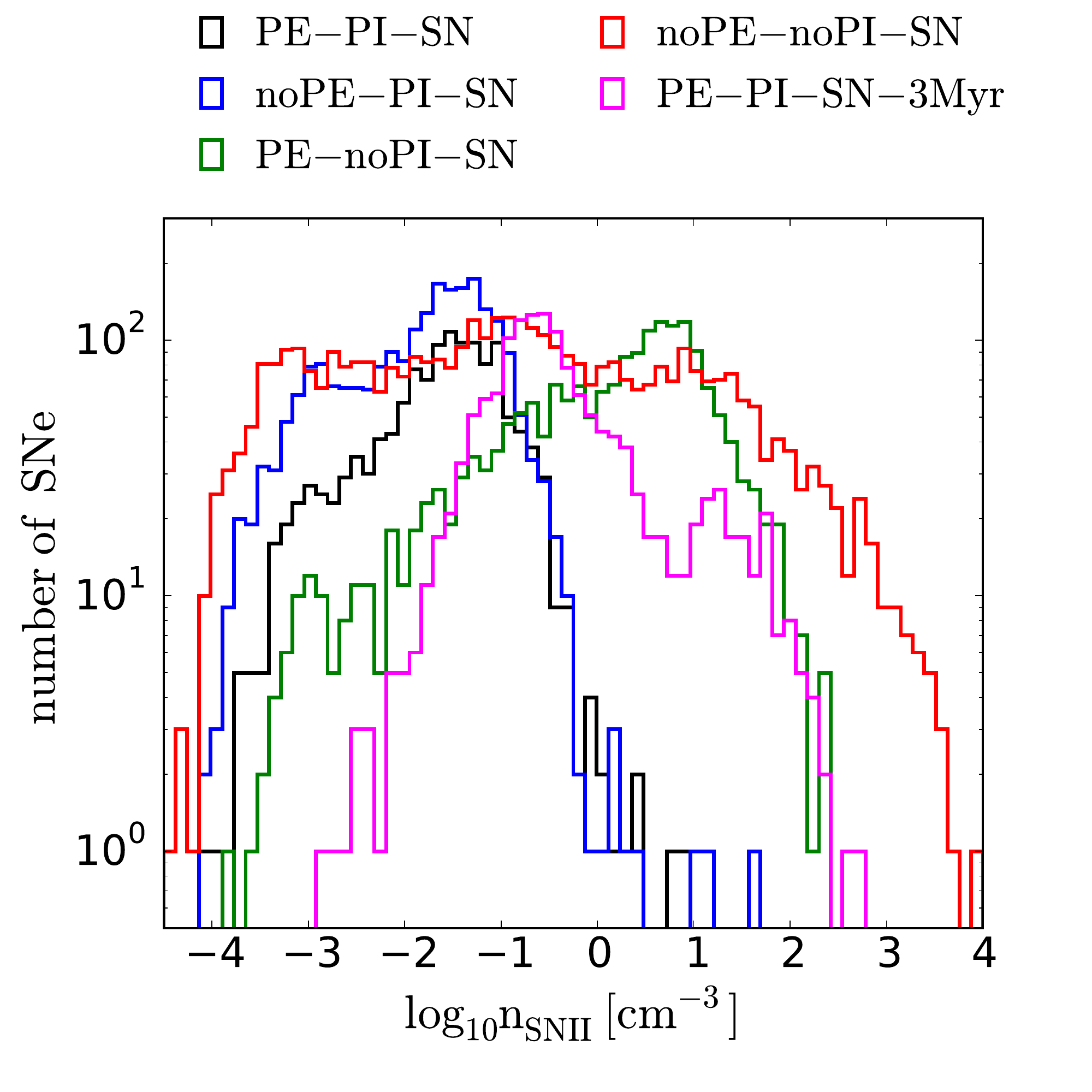}
	\caption{
		Histogram of the ambient hydrogen number density where the SNe occur ($n_{\rm SNII}$). 
		The more SNe that occur at low densities, the stronger the outflow rate is (cf. Fig. \ref{fig:outflow_mass_loading_2kpc_time}). 
		SNe occurring at high densities are ``futile'' events which show no correlation to the outflow rate.
		Photoionization can have both positive and negative effect on the outflow rate,
		as it not only reduces $n_{\rm SNII}$ (positive) but also reduces the amount of star-forming gas and thus the SFR (negative). 
		Adopting a too short stellar lifetime (3 Myr) suppresses outflows, as it gives SN progenitors too little time to escape from their parent clouds.}
	\label{fig:hist_rhoSNII}
\end{figure}

Run \textit{PE-PI-SN} has a higher outflow rate than run \textit{PE-noPI-SN}, 
which suggests that photoionization can help drive outflows by reducing the ambient density at the locations where SNe will later explode (see Eq. 10 in \citealp{2016arXiv161206891N}).
This can be seen in Fig. \ref{fig:hist_rhoSNII}, 
which shows a histogram of the ambient number density where the supernovae have gone off ($n_{\rm SNII}$) before $t$ = 800 Myr in the same set of runs.

Turning off PE heating (\textit{noPE-PI-SN} and \textit{noPE-noPI-SN}) increases the outflow rate because of the increased SFR.
The gas in the cold phase becomes colder without the PE heating (cf. Fig. \ref{fig:PD_3by2}) which makes the gas more gravitationally unstable. It is therefore
easier for it to collapse to high densities, resulting in more star-forming gas (cf. Fig. \ref{fig:h2_vs_time}, bottom panel).
The effect of photoionization in the two \textit{noPE} runs turns out to be the opposite here,
where \textit{noPE-noPI-SN} has a slightly higher outflow rate.
This is because not only does photoionization reduce $n_{\rm SNII}$, which promotes outflows,
it can also suppress star formation and therefore suppress outflows.
The net effect of photoionization on the outflow rate therefore depends on the competition between these two factors.

Run \textit{PE-PI-SN-3Myr} has the lowest mass loading factor, with an average value of 0.1.
Compared to run \textit{PE-PI-SN}, which has a mass loading factor $\sim$ 1,
it demonstrates that the adopted stellar lifetime can have a profound effect on the outflows,
even though it seems to only have a very weak effect on the SFR (cf.\ Fig. \ref{fig:sfr_vs_time}).
This is because 3 Myr is much shorter than the lifetime of the majority of the massive stars, some of which can have lifetimes 
as long as 30 Myr. Adopting an artificially short stellar lifetime has two consequences.
First, photoionization will have less time to operate and thus will be less efficient at reducing $n_{\rm SNII}$. 
However, the largest ionizing photon luminosities come from the most massive stars, which have lifetimes comparable to
3~Myr, and so it is unclear how important this effect is in practice.
Secondly, and perhaps more importantly,
the supernova progenitors that have small ionizing photon luminosities will have less time to drift away from their parental clouds.
As a result, these SNe will mostly explode within high density environments.

\begin{figure}
	\centering
	\includegraphics[trim = 0mm 10mm 0mm 30mm, clip, width=1\linewidth]{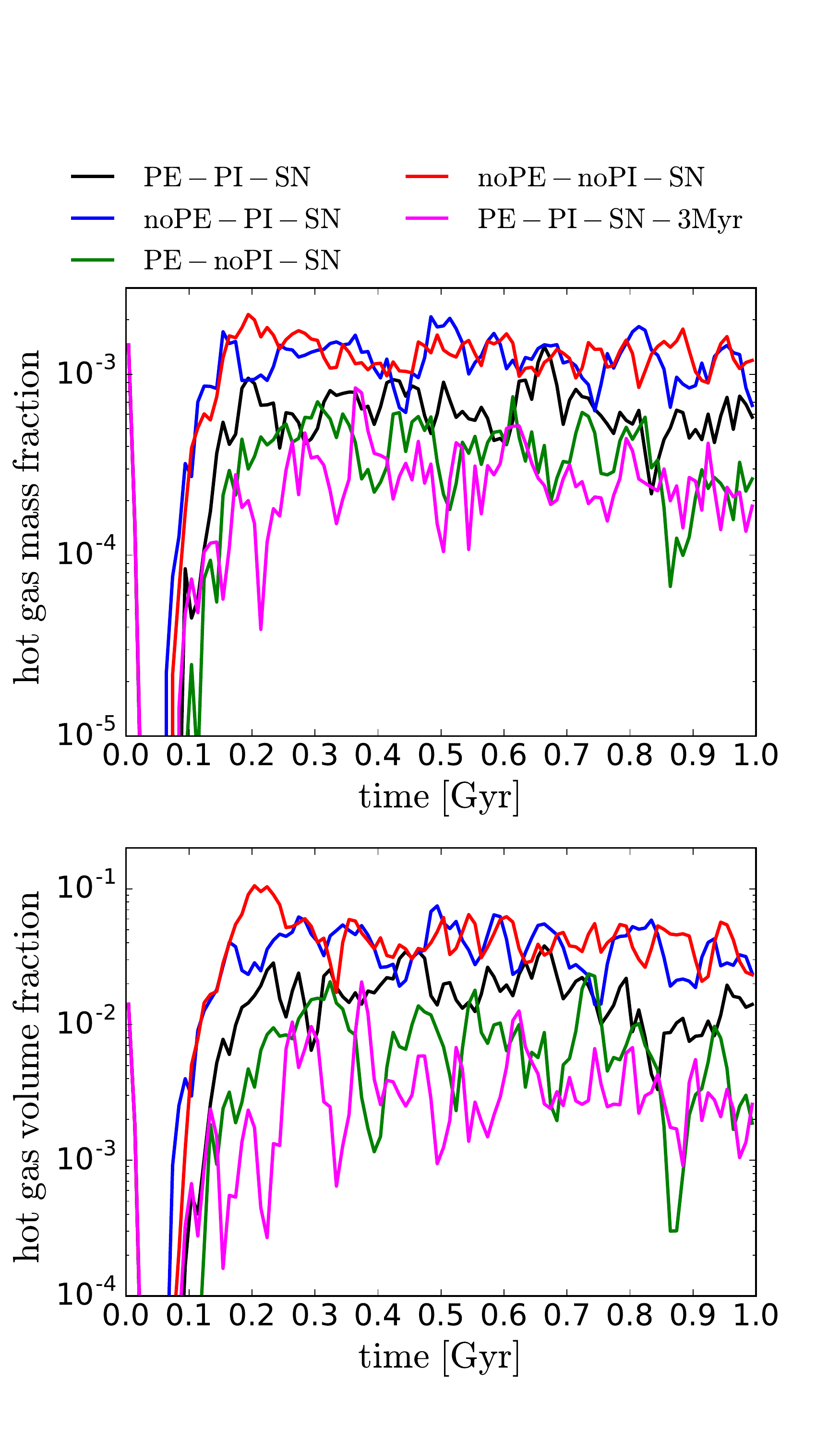}
	\caption{
		Time evolution of the hot gas fraction in mass (upper panel) and in volume (lower panel) in the region of $R <$ 1.5 kpc and $|z| < $ 0.2 kpc. 
		The outflow rate (cf. Fig. \ref{fig:outflow_mass_loading_2kpc_time}) increases with the hot gas fraction, 
		as the outflows are driven by the over-pressurized hot gas.}
	\label{fig:hot_MVfrac}
\end{figure}

It is worth noting that the ranking in outflow rate (Fig. \ref{fig:outflow_mass_loading_2kpc_time}) among different runs is the same as that in Fig. \ref{fig:hist_rhoSNII} in the low-density regime ($n_{\rm SNII} \lesssim 0.1$ cm$^{-3}$).
On the other hand, 
SNe going off in high-density environments seem to have little impact on the outflow rate. They can be viewed as ``futile events'',
as most of their injected energy will be radiated away rapidly.
Photoionization turns most of the futile SN events into ``efficient'' events,
but it does not necessarily guarantee higher outflow rate as it can also reduce the total number of SN events by suppressing star formation.
Eventually, it is the number of efficient SN events (instead of the fraction of the total SN events that are efficient) that controls the outflow rate.

Fig. \ref{fig:hot_MVfrac} shows the time evolution of the hot gas ($T > 3\times 10^4$ K) fraction both in mass (upper panel) and in volume (lower panel) in the region of $R <$ 1.5 kpc and $|z| < $ 0.2 kpc.
The ranking of different runs is the same as that in the outflow rates (Fig. \ref{fig:outflow_mass_loading_2kpc_time}) and the number of low-$n_{\rm SNII}$ SN events (Fig. \ref{fig:hist_rhoSNII}).
The implication is clear:
when there are more SNe occurring in low density environments,
either because of photoionization prior to the SN events or because the progenitor stars drift away from their parental clouds,
these supernova remnants can retain their thermal energy for a longer time (as the radiative cooling rate scales with the square of the density), 
leading to a higher hot gas fraction that provides the necessary pressure force to drive the outflows.
This picture is also qualitatively consistent with the gas morphology shown in Fig. \ref{fig:faceOnMaps500Myr} and \ref{fig:edgeOnMaps500Myr}:
\textit{PE-PI-SN-3Myr} shows the smallest bubbles, and it has the lowest hot gas fraction, the fewest efficient SN events, and the lowest outflow rate.

\section{Discussion}\label{sec:discussion}

\subsection{The ``primary'' feedback process}

Given the different combinations of feedback we have explored,
it is tempting to single out a feedback process that is the ``most important''.
This may seem to be impractical due to the potential nonlinear coupling among different feedback processes \citep{2014MNRAS.445..581H}.
Nevertheless,
it is still interesting to ask which process by itself can lead to a galaxy closest to the full-feedback model.
Because of the simplistic treatment we adopt for photoionization,
we will just focus on PE heating and SN feedback.

Judging from the total SFR (Fig. \ref{fig:sfr_vs_time} and \ref{fig:ksplot}), 
at low enough surface densities (as in our fiducial disk),
PE heating alone suppresses star formation slightly more  efficiently than SNe alone (comparing \textit{PE-noPI-noSN} and \textit{noPE-noPI-SN}).
However,
the difference is only around a factor of two.
It is therefore insufficient to assess the relatively importance of these two processes in suppressing star formation merely based on the total SFR.

On the other hand,
the gas morphology and the phase diagram depend much more crucially on SNe than on PE heating (Fig. \ref{fig:faceOnMaps500Myr}, \ref{fig:edgeOnMaps500Myr} and \ref{fig:PD_3by2}).
If we assume that the full-feedback run \textit{PE-PI-SN} gives the most ``correct'' solution, then 
one can still obtain gas morphologies and phase distributions with SNe alone (\textit{noPE-noPI-SN}) that are close to the correct ones. This is because the SNe provide much more heating in the diffuse phase than the PE heating.
Though one does end up with slightly more cold and dense gas due to the lack of PE heating,
the way star formation is suppressed is qualitatively the same:
SNe regulate the amount of cold and dense gas that can form out of the warm and diffuse gas reservoir.

However, the morphology and phase diagram with PE heating alone (\textit{PE-noPI-noSN}) differ substantially from the other runs.
For the dense gas, PE heating is a mild process that only heats up the gas to a few hundred Kelvin at most
and thus cannot inject enough energy and momentum into the ISM to disperse the clouds or drive turbulence.
As such, the majority of gas remains cold and simply forms dense gas clumps due to gravity. When the gas surface density is low, these clumps can reach hydrostatic equilibrium and stop further collapse due to the locally enhanced PE heating. 
The way star formation is suppressed here is therefore different even qualitatively from that in the SNe runs.
Moreover, this PE heating is artificially boosted, as in the ``real'' ISM (\textit{PE-PI-SN}) the gas is on average much farther away from the stars (see Fig. \ref{fig:rho_vs_G0}). This is a perfect example of how different feedback processes couple with each other non-linearly.

It is worth noting that the effect of SNe (or any other feedback processes), when operating alone, could also be artificially enhanced,
though to a lesser extent as we argue above.
An additional and perhaps more compelling piece of evidence indicating the dominant importance of SNe in regulating star formation comes from the fact that the majority of gas in \textit{PE-PI-SN} is much warmer than its thermal equilibrium temperature (as shown in the one-zone model in Fig. \ref{fig:PD_eq_5}),
which suggests that SN heating dominates over PE heating.

These considerations, together with the fact that the PE-only run does not drive outflows at all, lead us to conclude that SN feedback is the primary process responsible for shaping the ISM and regulating star formation in dwarf galaxies, with PE heating playing a secondary (but still important) role.


\subsubsection{Comparison with  \citet{2016Natur.535..523F}}
Our results are in conflict with those obtained by \citet{2016Natur.535..523F}, who also studied the relative
importance of SN feedback and PE heating for regulating star formation in a low metallicity, low surface density dwarf galaxy.
\citet{2016Natur.535..523F} carried out their simulations using the {\sc enzo} hydrodynamical code \citep{2014ApJS..211...19B} and used a modified
version of the {\sc grackle} cooling and chemistry package \citep{2016arXiv161009591S} to model the thermal evolution of the ISM. 
They found, using a similar setup to that studied in this paper, that SNe alone were ineffective at suppressing star formation, but that if they also
included PE heating by a spatially and temporally varying ISRF, the resulting star formation rates were in good agreement with observation. They
also found that they obtained very similar star formation rates in runs with only PE heating to those in runs with both PE heating and SN feedback,
leading them to conclude that PE heating was the dominant feedback process. Finally, they found a large amount of cold gas in all of their runs,
while in our runs, the ISM is always dominated by warm gas whenever SNe are included.

In our present study, we have adopted a self-consistent treatment of a spatially and temporally varying ISRF very similar to that used in
\citet{2016Natur.535..523F}, so it seems unlikely that our treatment of the PE heating is responsible for the tension in our results. Instead,
we believe that the primary cause of this tension is an inconsistency in the Forbes et~al. thermal model. \citet{2016Natur.535..523F} model
the effects of metal heating and cooling in the ISM using a cooling table derived from the {\sc Cloudy} photodissociation region code 
\citep{2013RMxAA..49..137F}.
Specifically, they use a table that was derived using the method of \citet{2008MNRAS.385.1443S} and that assumes an 
incident UV radiation field given by the $z = 0$ model from \citet{2012ApJ...746..125H}.\footnote{This is the file \texttt{CloudyData\_UVB=HM2012.h5} in the
{\sc grackle} package.} However, in an attempt to account for the effects of atomic hydrogen 
self-shielding in dense gas, they modified the way that this table was used within {\sc grackle}.\footnote{See specifically change set  3a58638 in
their fork of the {\sc grackle} codebase, available at \texttt{https://bitbucket.org/jforbes/grackle}}
They switched off the heating term in gas denser than $n_{\rm gas} > 0.01$ cm$^{-3}$, reasoning that this gas should be self-shielded and hence should not
be strongly heated by the extragalactic UV background (J.~Forbes, private communication).

Unfortunately, doing this without modifying the cooling rates leads to an inconsistency. The cooling rate in gas with $T < 10^{4}$~K is
highly sensitive to the fractional ionization \citep[see e.g.\ Fig.\ 2 in][]{1972ARA&A..10..375D}, both because the fine structure cooling
rates are sensitive to the free electron number density, and also because hydrogen recombination cooling can be an important process 
in this regime. The cooling rates computed using {\sc cloudy} and tabulated in the file \texttt{CloudyData\_UVB=HM2012.h5} were computed
under the assumption that the gas was optically thin to the \citet{2012ApJ...746..125H} UV background, i.e.\ they do not account for the
effects of self-shielding. Therefore, the modified version of {\sc grackle} used by \citet{2016Natur.535..523F} calculates the heating rate of
$n_{\rm gas} > 0.01$ cm$^{-3}$ gas under the assumption that it is self-shielded but calculates the cooling rate for the same gas under the
assumption that it is optically thin, which is inconsistent. The consequence of this is that in gas with $n_{\rm gas} > 0.01$ cm$^{-3}$, the
calculations carried out by \citet{2016Natur.535..523F} implicitly assume a fractional ionization that is much larger than it should be in 
self-shielded gas, and hence substantially overestimate the cooling rate, by an order of magnitude or more. This overestimated cooling rate 
leads to the production of large quantities of cold
gas in their simulations, and consequently to an extremely large star formation rate in the absence of feedback. Because the gas cools
so quickly, SN feedback is unable to strongly suppress this rapid star formation (primarily because it acts too late), and so a reasonable 
SFR is obtained only once one includes the effects of photoelectric heating. 

Simulations using a similar setup to that in \citet{2016Natur.535..523F} but that
correct the cooling inconsistency (by switching to a cooling
table appropriate for the case of a zero UV background in gas with $n_{\rm gas} > 0.01$ cm$^{-3}$) 
amongst other changes (such as introducing some initial turbulence to prevent over-rapid settling of the gas disk
and adopting a smaller gas scale length)
find results which differ significantly
from those in the previous simulations (J.~Forbes, private communication). 
The large amounts of cold, low-density gas vanish and the
ISM becomes dominated by warm gas, in good agreement with our results. In addition, the star formation rate in the absence of feedback
decreases significantly, as does the star formation rate with only SN feedback. Forbes et~al.\ continue to find that runs with PE heating alone
or with PE heating plus SN feedback are more effective at suppressing the star formation rate than pure SN feedback (i.e.\ the results of
\citealt{2016Natur.535..523F} remain qualitatively unchanged), but the difference in the star formation rate between the SNe only and
SNe+PE runs is now only a factor of two to three, rather than the factor of 20 found in the previous simulations (J.~Forbes, private
communication). This significantly alleviates  the tension with our results. However, it does not eliminate it entirely: we find that the SFR
in roughly comparable in runs with SNe but without PE heating and with PE heating but without SNe (runs \textit{noPE-PI-SN} and \textit{PE-noPI-noSN}; see Fig.\ \ref{fig:sfr_vs_time})
whereas Forbes et~al.\ continue to find more effective suppression of star formation in their PE-only model than in their SN-only model.\footnote{It should be noted that the ``SN-only'' model in \citet{2016Natur.535..523F} actually includes SNe, photoionization and stellar winds from massive stars, with these three processes always being switched on and off together.
For this reason, our \textit{noPE-PI-SN} run is the best analogue of their SN-only run. We do not include feedback from stellar winds in our model, but we expect its effect should be sub-dominant at the metallicity we adopt (see Section \ref{sec:winds}).}
However, as we argue above,
this does not necessarily mean that PE heating is more effective than SNe in dwarf galaxies as the former can be artificially boosted in a PE-only run.

\subsection{Potential Caveats}

\subsubsection{Cosmic ray ionization}

For simplicity and to aid comparison with \citet{2016Natur.535..523F},
we have deliberately neglected the heating from cosmic ray (CR) ionization which was included in Paper I.
CR ionization heating is less important in the dense phase of the ISM compared to PE heating,
while it can dominate over PE heating in the diffuse phase.
However, 
as shown in Paper I, 
SN feedback still dominates when a constant CR ionization rate $\xi = 3\times 10^{-18}$ s$^{-1}$ is adopted,
where $\xi$ is scaled with the star formation rate surface density according to the Milky Way value. 
Therefore, we do not expect our results to change dramatically if CR ionization is included with a reasonable choice of $\xi $.

It is desirable to model the CR ionization heating self-consistently in a similar spirit to our treatment of PE heating.
However, this is much more challenging owing to the complicated behavior of CR propagation which depends on the configuration of the magnetic fields and some uncertainties in its propagation mechanisms \citep{2012MNRAS.423.2374U, 2013ApJ...777L..38H,  2016ApJ...816L..19G, 2016ApJ...827L..29S}.
A fully self-consistent treatment of the CR ionization heating therefore requires substantial further investigation.

\subsubsection{Treatment of photoionization}

As mentioned in Section \ref{sec:PI},
our treatment of photoionization suffers from mass biasing which preferentially ionizes mass concentrations as it has no angular resolution.
Therefore,
we expect it to over-reduce the amount of high-density gas and thus over-suppress star formation and over-promote galactic outflows.
Since in our simulations the SFR is insensitive to the inclusion of photoionization (see Fig. \ref{fig:sfr_vs_time}),
the overestimation should be rather small.
However,
the outflow rate does show notable dependence on photoionization (e.g., compare \textit{PE-PI-SN} and \textit{PE-noPI-SN} in Fig. \ref{fig:outflow_mass_loading_2kpc_time}),
which can potentially be of numerical origin due to our simplistic treatment.
The amount of overestimation will depend crucially on the density structure in the vicinity of photoionizing massive stars.
High-resolution simulations coupled with radiative transfer will provide important information on the credibility of this simplified but cost-efficient approximation.

\subsubsection{Stellar winds}\label{sec:winds}

Another important feedback mechanism we have ignored is the stellar winds from massive stars,
which can play a similar role as photoionization in the sense that they both can create favorable low-density environments for the subsequent SNe to be efficient and thus promote outflows \citep{2016arXiv160605346G,2016arXiv161006569P}. We have neglected them in this study for two reasons, one numerical and one physical. Numerically, the issue is that it is difficult to properly model stellar winds using SPH owing to the fixed SPH mass resolution. This limits the ability of SPH to capture the low but continuous mass-loss rate from massive stars, unless the mass resolution is very much better than in our whole galaxy simulations \citep[see e.g.][]{2013MNRAS.436.3430D}. In other words, it is much more difficult to obtain a resolved and converged evolution of stellar winds using SPH as opposed to in an Eulerian grid-based code that in principle has infinite mass resolution (albeit coupled with finite spatial resolution). 

Physically, the justification for ignoring stellar winds in low-metallicity dwarf galaxies stems from the fact that the mass loss rates of OB stars and Wolf-Rayet stars are highly sensitive to metallicity \citep{2000A&AS..141...23P,2005A&A...442..587V}. Reducing the metallicity reduces the radiative forcing and hence greatly reduces the strength of the winds. Consequently, at the metallicity examined in this study ($Z = 0.1 \, Z_{\odot}$), stellar winds will be much less effective and much less energetic than in the local ISM. We therefore do not expect them to be a significant source of feedback in comparison to photoionization, justifying our decision to ignore them. 

\section{Summary}\label{sec:summary}

We present high resolution ($m_{\rm gas} = 4 {\rm M_\odot}$, $N_{\rm ngb}$ = 100) simulations of isolated gas-rich dwarf galaxies which include self-gravity, non-equilibrium cooling and chemistry, star formation, stellar feedback and metal enrichment. 
Each simulation is run for 1 Gyr.
We introduce a flexible method to sample the IMF which allows for the formation of individual massive stars at our resolution.
The stellar feedback includes PE heating, photoionization and SNe.
Instead of assuming a constant value of $G_0$ throughout the disk,
we self-consistently calculate $G_0$ from the star particles which makes it vary both spatially and temporally.
Photoionization is treated with a Str\"{o}mgren-volume approach that can account for overlapping H{\sc ii} regions. 
In a uniform medium, 
the dynamical evolution of the HII regions converges with increasing resolution, and agrees well
with that predicted by more sophisticated radiative transfer methods (Fig. \ref{fig:hiiconvergen100}),
though in an inhomogeneous medium the method becomes strongly mass-biased.
We run simulations with two initial conditions which only differ in the scale-length of the gas disk and therefore the central gas surface density ($\Sigma_{\rm gas}$).

We find that SNe play a dominant role in suppressing star formation and shaping the ISM in dwarf galaxies.
The run with SNe alone is able to reproduce most of the ISM properties seen in  the full-feedback run,
such as the star formation rate (Fig. \ref{fig:sfr_vs_time}), gas morphology (Fig. \ref{fig:faceOnMaps500Myr} and \ref{fig:edgeOnMaps500Myr}), gas distribution in the phase diagram (Fig. \ref{fig:PD_2by2}) and the galactic outflow rate (Fig. \ref{fig:outflow_mass_loading_2kpc_time}).
Because of SNe, 
the majority of gas (the diffuse phase) becomes much warmer than it should be if it were in thermal equilibrium.
In other words,
the dynamical effect of SNe results in a new thermal-equilibrium relation, 
which makes the gas lie above the equilibrium curves in the one-zone model as shown in Fig. \ref{fig:PD_eq_5}
where PE heating is the main heat source.
This demonstrates that SN heating largely dominates over PE heating in the ISM of dwarf galaxies,
and star formation is regulated by SNe which control the warm-to-cold transition of gas.
However,
as far as chemistry is concerned,
SNe are not an important destruction mechanism of H$_2$ compared to photodissociation,
as the SN-only run is not able to suppress the H$_2$ fraction to the same level as the full-feedback run (Fig. \ref{fig:h2_vs_time}).

On the other hand,
the run with PE heating alone, while being able to suppress star formation to a similar level as the full-feedback run,
generates drastically different results for all other physical properties.
More importantly,
the effect of PE heating is artificially boosted in the PE-only run where the dense clouds are always located very close to the UV-emitting stars due to gravity and the lack of strong feedback mechanism to disperse the clouds (Fig. \ref{fig:rho_vs_G0}).
As such,
the mechanism that suppresses star formation in the PE-only run does not exist in the realistic ISM of dwarf galaxies,
as it is not possible to have so much gas located so close to the UV sources without being dispersed by other feedback processes (such as SNe or photoionization).
This is a good example of the nonlinear coupling between different feedback processes.
The artificially boosted ISRF also leads to a significantly enhanced dust continuum IR emission (by an order of magnitude)
and therefore an FIR line-to-continuum ratio too low compared to the observed nearby dwarf galaxies (Fig. \ref{fig:compare_DGS}).
Including SNe leads to an FIR line-to-continuum ratio of the simulated galaxies that agrees well with observations by pushing gas further away from the massive stars.

That said,
the PE heating still has a notable (though secondary) effect on the ISM.
PE heating reduces the amount of cold and dense gas and its inclusion in the simulation reduces the SFR by around a factor of two compared to the result in a simulation with SNe but without PE heating. In addition, PE heating indirectly reduces the outflow rate by reducing the SN rate and the hot gas fraction.

In our compact disk setup,
neither SNe nor PE heating is able to suppress star formation by itself at a rate similar to that found in the full-feedback run (Fig. \ref{fig:sfr_vs_time}).
Including photoionization is crucial in this case,
which highlights the importance of pre-SN feedback from massive stars when the gas surface density becomes higher.

Photoionization can also affect the ISM in a similar way as PE heating by reducing the amount of star-forming gas and therefore the SFR.
However,
it does not necessarily reduce the outflow rate.
This is because unlike PE heating that can only mildly heat up the dense gas (to a few hundred of Kelvin at most, see Fig. \ref{fig:PD_eq_5}) and make it gravitational stable,
photoionization heats the dense gas to a much higher temperature ($10^4$ K) which can create expanding bubbles due to the large pressure imbalance.
Therefore,
it can prepare a suitable low-density environment for the subsequent SNe to be efficient and therefore promote outflows.
The net effect of photoionization on the outflows can thus be either positive or negative,
depending on the competition between the two processes.

We find clear evidence that the galactic outflow rate is dictated by the number of SNe that occur in low-density environments (Fig. \ref{fig:outflow_mass_loading_2kpc_time} and \ref{fig:hist_rhoSNII}),
while the number of SNe that occur in dense environments shows little correlation with the outflow rate (see also \citealp{2016arXiv161206891N}, Eq. 10).
The outflow rate also positively correlates with the hot gas fraction in the ISM (Fig. \ref{fig:hot_MVfrac}),
as it is the over-pressurized hot gas that drives the outflows.
Our result is in very good agreement with recent studies such as \citet{2015MNRAS.454..238W},  \citet{2016MNRAS.456.3432G} or \citet{2016arXiv160605346G}.
In addition to the effect of photoionization,
we find that the adopted stellar lifetime also has an important effect on $n_{\rm SNII}$.
Adopting a short stellar lifetime (3 Myr for all massive stars) reduces the time available for the SN progenitors to escape from their birth clouds,
which increases $n_{\rm SNII}$ and reduces the outflow rate.

\section*{Acknowledgments}
We thank John Forbes for valuable comments on the manuscript,
and we thank Bruce Elmegreen, Sambit Roychowdhury and Diane Cormier for providing the observational data.
The Center for Computational Astrophysics is supported by the Simons Foundation.
CYH, TN, SCOG and SW acknowledge support from the Deutsche Forschungsgemeinschaft  (DFG) priority program SPP 1573 ``Physics of the Interstellar Medium''.
SCOG acknowledges additional support from the DFG via SFB 881, ``The Milky Way System'' (sub-projects B1, B2 and B8), and from the European Research Council under the European Community's Seventh Framework Programme (FP7/2007-2013) via the ERC Advanced Grant STARLIGHT (project number 339177).
TN acknowledges support from the Cluster of Excellence ``Origin and structure of the Universe''.
PCC acknowledges support from the Science and Technology Facilities Council (under grant ST/N00706/1) and the European Community’s Horizon 2020 Programme H2020-COMPET-2015, through the StarFormMapper project (number 687528).
SW additionally thanks the DFG for support through SFB 956, ``Conditions and impact of star formation'' (sub-project C5), and the European Research Council under the European Community’s Eighth Framework Programme FP8 via the ERC-Starting-Grant RADFEEDBACK (proposal no. 679852).


\bibliographystyle{mn2e}
\bibliography{literatur}


\appendix

\section{Convergence test of the D-type expansion}\label{app:Spitzer}

In this test,
four star particles are placed the in the center of a uniform box which has a number density $n_{\rm gas}$ = 100 cm$^{-3}$ and gas temperature $T = 10^3$ K.\footnote{This ambient temperature of $10^3$ K is much higher than the typical temperature in the star-forming gas which usually has $T < 100$ K.
The value is chosen in order to more easily compare with results obtained from radiative transfer codes in \citet{2015MNRAS.453.1324B} (see below).}
The medium is composed of pure hydrogen,
i.e.\ $X_{\rm H} = 1$.
Each star particle has an ionizing photon rate of $2.5 \times 10^{48}$ s$^{-1}$ so that the total ionizing photon rate is $S_{\rm*,tot} = 10^{49}$ s$^{-1}$,
which is typical for an O star.
This leads to an initial Str\"{o}mgren radius $R_{\rm s}$ = 3.1 pc.

With this setup, we expect the ionizing photons from the central sources to create an H{\sc ii} region bounded by an ionization front at position $R_{\rm IF}$. Initially, when $R_{\rm IF} \ll R_{\rm s}$, the front is R-type and expands rapidly. However, once $R_{\rm IF}$ approaches $R_{\rm s}$, the expansion velocity of the front decreases rapidly and the front undergoes a transition to a D-type front. Our photoionization algorithm does not capture the initial evolution of the R-type ionization front, since we assume that the ionization front expands to $R_{\rm IF} = R_{\rm s}$ instantaneously. However, the code should be capable of following the subsequent D-type expansion of the front. Since this can be solved for analytically in this simple scenario, this provides a useful test of code and allows us to establish the mass resolution required to accurately follow the D-type expansion.

\begin{figure}
	\centering
	\includegraphics[width=0.9\linewidth]{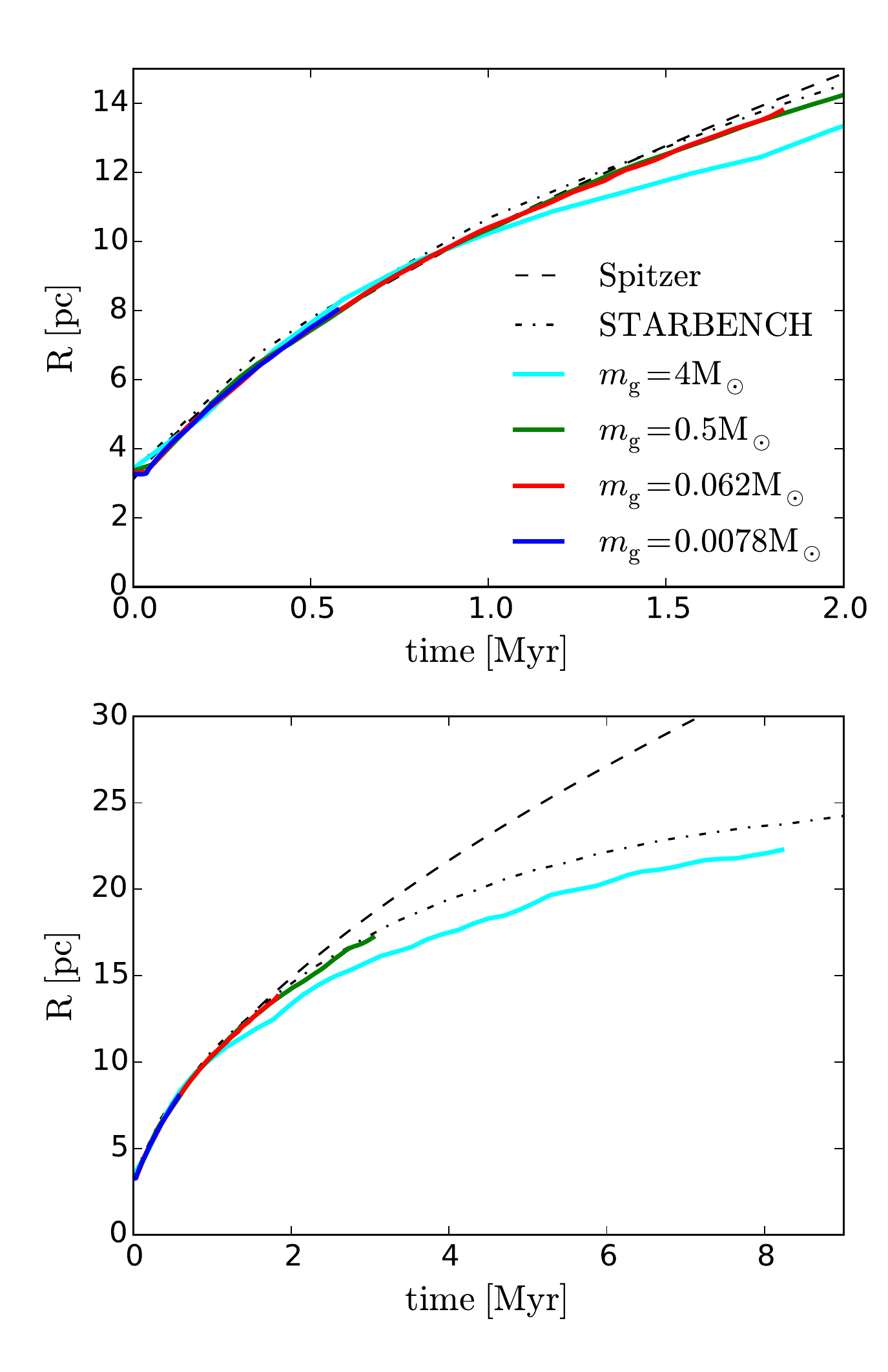}
	\caption{
		Time evolution of the ionization front $R_{\rm IF}$ produced by  
		four stars with an ionizing photon rate of $2.5 \times 10^{48}$ s$^{-1}$ each, located in the center of a uniform box with initial density $n_{\rm gas} = 100 \, {\rm cm^{-3}}$ and temperature $T = 10^3$ K. 
		The convergence test is performed with four different resolutions, with gas particle masses $m_{\rm g} = 4 {\rm M_\odot}, \, 0.5 {\rm M_\odot}, \, 0.625 {\rm M_\odot}$, and $0.0078125 {\rm M_\odot}$, respectively.
		The dashed line shows the Spitzer solution \citep{1978ppim.book.....S}, which accurately describes the early-time evolution, 
		while the dashed-dotted line shows the semi-empirical result obtained by the  {\sc starbench} project \citep{2015MNRAS.453.1324B}, which accurately describes the late-time 
		behaviour of the front. At early times, all of the runs agree well with the analytical prescription, while at late times, the results converge to the  {\sc starbench} solution as long as $m_{\rm g} \leq 0.5 {\rm M_\odot}$.}
	\label{fig:hiiconvergen100}
\end{figure}

Fig. \ref{fig:hiiconvergen100} shows the time evolution of the ionization front radius $R_{\rm IF}$, defined here as the maximum radius where a gas particle with an ionization fraction $x_{\rm H^+} > 0.95$ can be found.
The test is performed with four different resolutions with gas particle masses $m_{\rm g} = 4 {\rm M_\odot}, \, 0.5 {\rm M_\odot}, \, 0.625 {\rm M_\odot}$, and $0.0078125 {\rm M_\odot}$, respectively. The dashed line shows the classic Spitzer solution \citep{1978ppim.book.....S}
\begin{equation}
	R_{\rm Spi} = R_{\rm s} \left(1 + \frac{7}{4} \frac{c_i t}{R_{\rm s}} \right)^{4/7},
\end{equation}
where $c_i$ is the sound speed of the ionized medium and $t$ is the simulation time.
The Spitzer solution describes the early evolution of $R_{\rm IF}$ when the pressure imbalance across $R_{\rm IF}$ is large.
At a later stage, as the H{\sc ii} region expands,
the pressure imbalance decreases and eventually the expansion stops when it reaches pressure equilibrium,
as opposed to the Spitzer solution which expands forever. This late-time evolution is investigated in \citet{2015MNRAS.453.1324B} using several different radiative transfer codes as part of the code comparison project {\sc starbench}. This study found a simple semi-empirical prescription that accurately represented the late-time behaviour of the D-type front, which is shown as the dashed-dotted line in Fig. \ref{fig:hiiconvergen100}. The upper and lower panels in the Fig. show the early and late evolution, respectively.

As described in Section \ref{sec:PI},
our approach is able to account for overlapping HII regions.
This can be seen at $t$ = 0 in Fig. \ref{fig:hiiconvergen100} where all four runs accurately reproduce $R_{\rm s}$ = 3.1 pc.
This is different from the non-iterative approach sometimes adopted in the literature, where overlapping HII regions suffer from ``double-counting''. For this test problem, this non-iterative approach would lead to four times fewer gas particles becoming ionized than should be the case.  
Moreover,
our neighbor search is done globally instead of only on the local domain and thus is independent of domain decomposition\footnote{In this test we have run on both 16 and 32 processors in parallel and found essentially the same results}.

In the early stage, as expected, the Spitzer solution and the {\sc starbench} solution agree well with each other.
Our four runs show excellent converging behavior which agrees with the analytic solution at all resolutions until $t$ = 1 Myr,
when the $m_{\rm g} = 4 {\rm M_\odot}$ run starts to underestimate $R_{\rm IF}$.
At $t \gtrsim$ 2 Myr,
the Spitzer solution starts to over-predict $R_{\rm IF}$ as it neglects the pressure of the ambient medium.
Except for the $m_{\rm g} = 4 {\rm M_\odot}$ run,
the other three runs show excellent agreement with the {\sc starbench} solution throughout the simulation time.
We conclude that with the resolution we adopted in this work ($m_{\rm g} = 4 {\rm M_\odot}$),
we expect to slightly underestimate $R_{\rm IF}$ in a uniform medium due to insufficient resolution, although even at late times the error in $R_{\rm IF}$ remains relatively small.

\section{Parameter study for the star formation threshold}\label{app:MJtest}

In this section we investigate the consequences of varying the star formation threshold $N_{\rm th}$ (defined in Section \ref{sec:SF_th}).
We run two additional runs that are identical to the \textit{PE-PI-SN} run except for the adopted $N_{\rm th}$:
\textit{PE-PI-SN-MJ2} and \textit{PE-PI-SN-MJ32} with $N_{\rm th}$ = 2 and 32, respectively; recall that in run \textit{PE-PI-SN},
$N_{\rm th} = 8$. Note that a smaller $N_{\rm th}$ corresponds to, effectively, a higher density threshold at a fixed temperature.

\begin{figure}
	\centering
	\includegraphics[trim = 0mm 30mm 10mm 10mm, clip, width=0.99\linewidth]{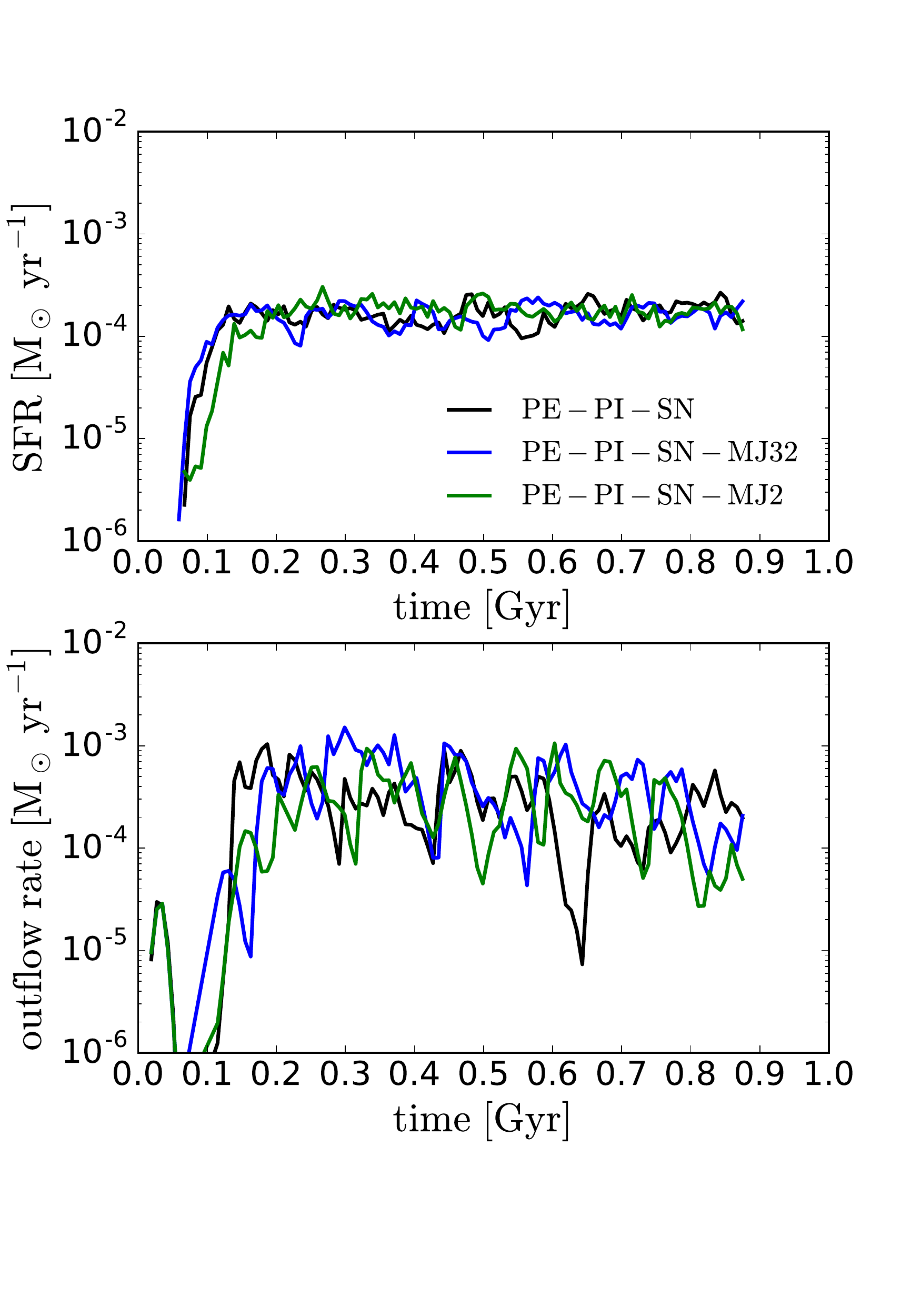}
	\caption{
		Time evolution of the SFR (upper panel) and the galactic outflow rate (lower panel) of the entire disk.
		\textit{PE-PI-SN-MJ2},\textit{PE-PI-SN}, \textit{PE-PI-SN-MJ32} adopt $N_{\rm th}$ = 2, 8 and 32, respectively.
		The SFR and the outflow rate seem to be insensitive to $N_{\rm th}$.
	}
	\label{fig:sfr_ofr_time_bin10Myr_MJ}
\end{figure}

Fig. \ref{fig:sfr_ofr_time_bin10Myr_MJ} shows the time evolution of the SFR (upper panel) and the galactic outflow rate (lower panel).
Run \textit{PE-PI-SN-MJ2} begins to form stars the latest (after $t \sim$ 100 Myr) as the gas has to collapse to higher densities and lower temperatures in order to be star-forming.
Besides that,
no systematic difference can be found among the three runs.
The SFR and the outflow rate seem to be insensitive to $N_{\rm th}$.

\begin{figure}
	\centering
	\includegraphics[trim = 20mm 10mm 10mm 0mm, clip, width=0.99\linewidth]{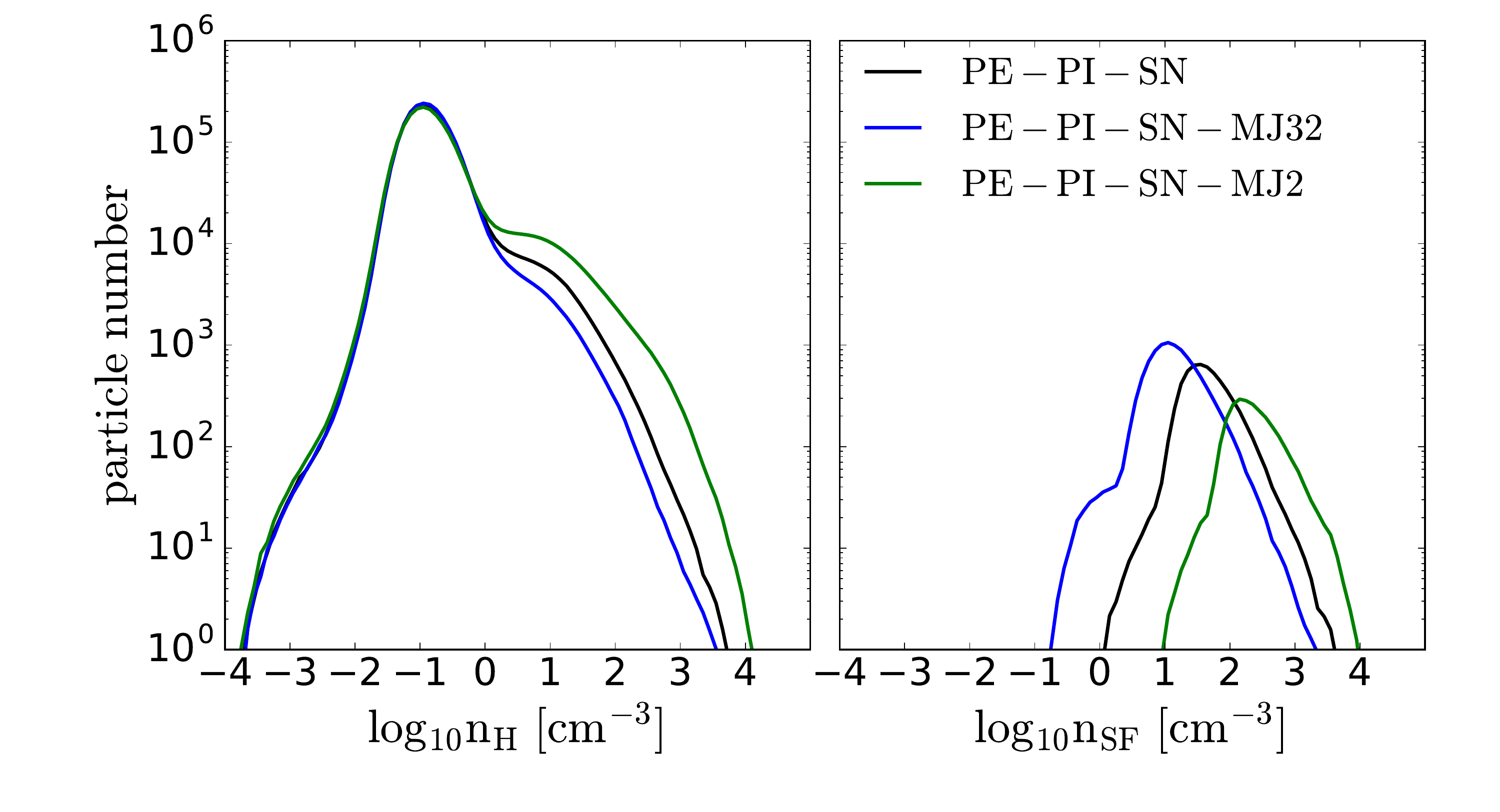}
	\caption{
		Histograms of the number density of the total gas (left panel) and the star-forming gas (right panel), 
		averaged over a period from $t$ = 400 Myr to $t$ = 900 Myr with a time interval of 10 Myr. 
		Smaller $N_{\rm th}$ leads to more dense gas as the gas has to collapse to higher densities and lower temperatures in order to form stars,
		which delays the subsequent stellar feedback to counter further collapse.
		On the other hand,
		the low density tail which corresponds to the hot bubble gas is insensitive to $N_{\rm th}$. 
	}
	\label{fig:rhoPDF_MJ}
\end{figure}

Fig. \ref{fig:rhoPDF_MJ} shows the histograms of the number density of the total gas (left panel) and the star-forming gas (right panel), 
averaged over a period from $t$ = 400 Myr to $t$ = 900 Myr with a time interval of 10 Myr. 
Smaller $N_{\rm th}$ leads to more dense gas as the gas has to collapse to higher densities and lower temperatures before it can form stars,
which delays the subsequent stellar feedback to counter further collapse.
The peak density of the star-forming gas also increases as $N_{\rm th}$ decreases for the same reason.
On the other hand, the low density tail which corresponds to the hot bubble gas is insensitive to $N_{\rm th}$. 

\begin{figure}
	\centering
	\includegraphics[width=0.99\linewidth]{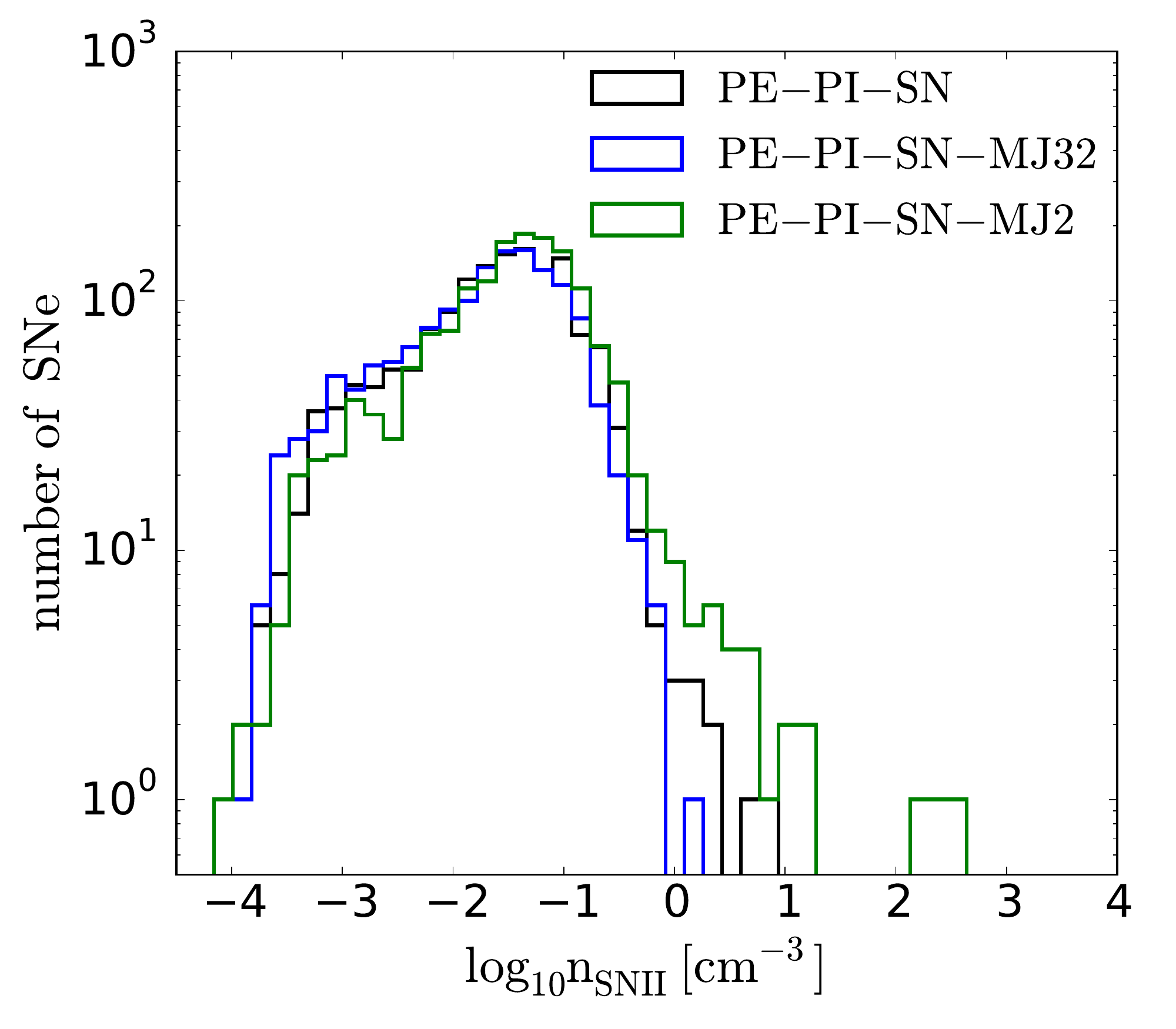}
	\caption{
		Histogram of the ambient hydrogen number density where the SNe occur ($n_{\rm SNII}$) before $t$ = 800 Myr. 
		Lowering $N_{\rm th}$ increases the number of SNe that occur at high densities, 
		while the number of SNe that occur at low densities remains unchanged.
		As a result, the outflow rate is also insensitive to $N_{\rm th}$ (see Fig. \ref{fig:sfr_ofr_time_bin10Myr_MJ}).
	}
	\label{fig:rhosniimj}
\end{figure}

Fig. \ref{fig:rhosniimj} shows the histogram of the ambient hydrogen number density where the SNe occur ($n_{\rm SNII}$) before $t$ = 800 Myr. 
Lowering $N_{\rm th}$ increases the number of SNe that occur at high densities, 
which would not affect the outflows as these are the futile events.
On the other hand,
the number of SNe that occur at low densities is insensitive to $N_{\rm th}$.
As a result, the outflow rate is also insensitive to $N_{\rm th}$ (see Fig. \ref{fig:sfr_ofr_time_bin10Myr_MJ}).

\end{document}